\documentclass{article}

\usepackage{arxiv}

\usepackage[utf8]{inputenc} 
\usepackage[T1]{fontenc}    
\usepackage{hyperref}       
\usepackage{url}            
\usepackage{booktabs}       
\usepackage{amsfonts}       
\usepackage{nicefrac}       
\usepackage{lipsum}
\usepackage{graphicx}
\usepackage{amsmath} 
\usepackage{setspace}
\usepackage{ulem}
\usepackage{algorithm}
\usepackage{algpseudocode}
\usepackage{xcolor,subfig,lipsum}
\usepackage{natbib}
\usepackage{rotating}
\graphicspath{ {./images/} }

\DeclareMathOperator{\logit}{logit}
\DeclareMathOperator{\expit}{expit}

\title{A multiple imputation approach to distinguish curative from life-prolonging effects in the presence of missing covariates}

\author{
 Marta Cipriani \\
  Mathematical Institute, Leiden University, Netherlands \\
  Department of Statistical Science, Sapienza University of Rome, Italy \\
  \texttt{m.cipriani@math.leidenuniv.nl} \\
   \And
 Marta Fiocco \\
  Mathematical Institute, Leiden University, Netherlands \\
  Department of Biomedical Data Science, Leiden University Medical Centre, Netherlands\\
  Trial and Data Center, Princess Maxima Centre for Pediatric Oncology, Netherlands \\
  \And
 Marco Alfò\\
  Department of Statistical Science, Sapienza University of Rome, Italy \\
  \And 
 Maria Quelhas\\
  \And
 Eni Musta\\
  Korteweg-de Vries Institute for Mathematics, University of Amsterdam, Netherlands \\
}

\begin{document}
\maketitle
\begin{abstract}
Medical advances have increased cancer survival rates and the possibility of finding a cure. Hence, it is crucial to evaluate the impact of treatments both in terms of cure and prolongation of survival. To achieve this, we may use a Cox proportional hazards (PH) cure model. However, a significant challenge in applying such a model is the potential presence of partially observed covariates. We aim to refine the methods for imputing partially observed covariates based on multiple imputation and fully conditional specification (FCS) approaches. To be more specific, we consider a general case in which different covariate vectors are used to model the probability of cure and the survival of patients who are not cured. We investigated the performance of the multiple imputation procedure based on the exact conditional distribution and an approximate imputation model, which helps to draw imputed values at a lower computational cost. To assess the effectiveness of these approaches, we compare them with a complete case analysis and an analysis that includes all available covariates in modelling both cure probabilities and the survival of the uncured. 
We discuss the application of these techniques to a real-world dataset from the BO06 clinical trial on osteosarcoma.
\end{abstract}

\keywords{mixture cure models \and multiple imputation \and chained equations \and osteosarcoma}

\newpage

\section{Introduction}
\label{section:intro}
Recent advances in medicine have increased the chances of survival for some types of cancer, such as breast cancer, melanoma, osteosarcoma, and acute childhood lymphocytic leukemia, and have even made cure possible. As a result, it is important not only to estimate the impact that several risk factors may have on survival, but also to analyse their effect on the probability of being cured. This is especially relevant in paediatric cancer. Osteosarcoma, a type of bone tumour that primarily affects children and young adults, may help illustrate this need. For this type of cancer, more than 95\% of patients who do not experience a relapse within 5 years after treatment are considered cured and are unlikely to relapse later \citep{ferrari2006late}. In survival analysis, individuals who will never experience the event of interest, such as disease recurrence or disease-related death, are referred to as "cured". Hence, it is crucial to consider cured patients in statistical analyses to assess the efficacy of a treatment in terms of cure rather than only focussing on its impact on prolonging survival.

Traditional survival analysis methods can be less informative when a portion of the population is considered cured, as they may not account for differences in cure status within the population. Furthermore, the cure status is not always observable; while we know that patients who experience the event of interest are uncured, we cannot determine whether censored patients will eventually experience the event. This led to the development of cure models, which provide a more suitable modelling approach, particularly in oncological studies \citep{legrand2019cure}.

One type of cure model is the mixture cure model, which defines overall survival as a combination of the probability of cure (incidence) and the survival of uncured patients (latency). Mixture cure models allow for the use of different covariates when modelling the incidence and latency components. This means that we can differentiate between the effect of a covariate on the probability of being cured and its effect on the survival of uncured patients. In this context, we consider a Cox proportional hazards (PH) cure model, where the incidence is modelled by a logistic regression model and a Cox PH model is used for the latency. Estimation is generally performed using the maximum likelihood approach in an incomplete data setting. For this purpose, the expectation-maximization (EM) algorithm is often used due to the latent cure status \citep{cai2012smcure}. For more detailed information on cure models, see the review paper by \cite{amico2018cure} or the book by \cite{book}. 

A recent study by \cite{musta2021new} examined the prognostic importance of the histological response and intensified chemotherapy on the cure rate and progression-free survival in a sample of osteosarcoma patients who have not been cured. However, the study excluded observations with missing histological responses from the analysis. This motivated us to investigate, in this article, the sensitivity of the results to the method used to handle missing values in the histological response.

When working with an incomplete dataset, a simple but naive approach to handling missing data is to conduct a complete case analysis, which involves removing observations with missing data. However, this method can lead to inefficient and inconsistent results when the missing mechanism is not completely at random \citep{little}. Alternatively, missing values can be filled in by using simple imputation techniques, either based on summaries (e.g. mean) or statistical models (e.g. regression) based on available data. While these methods allow for the use of the whole sample, completed by imputation, they may underestimate the final model's standard errors \citep{van2018flexible}. To address these issues, \cite{rubin1987multiple} introduced multiple imputation techniques. This approach produces unbiased estimates and incorporates the uncertainty of imputation into standard error estimates. Multiple imputation involves creating multiple imputed datasets, estimating the parameters in each of them separately, and then combining the results. When handling survival data, it has been found beneficial to include the survival outcome in the imputation model for the covariates \citep{white2009imputing}.

However, dealing with cured patients presents an additional challenge for the imputation procedure. \cite{beesley2016multiple} have developed a method for using multiple imputation in the Cox proportional hazards (PH) cure model, which is based on specifying either an exact or an approximate distribution of the missing covariates, conditional on the available ones and the survival response (time, status). The authors state in that paper that, for simplicity, they assume the same design vector is used in the models for both incidence and latency. This assumption restricts the applicability of the method in practice, as including all available covariates in both components of the mixture model may lead to the estimation of a large number of parameters. In addition, the factors affecting the probability of being cured can differ from those influencing the hazard, making the incidence and latency settings distinct from each other. Moreover, as demonstrated by \cite{ident}, the sharing of covariates between various components of a Cox PH cure model may prevent the model identifiability. Hence, a simpler model could be preferred (over the one with all covariates in both components) whenever supported by the data. 

In this paper, we consider the more general case of a mixture cure model with possibly different sets of covariates for the incidence and latency component. This allows us to fully leverage the benefits of a mixture model. Specifically, we consider three cases depending on whether the variable with missing values affects only the probability of cure, only the survival of the uncured, or both components. Building on the work of \cite{beesley2016multiple}, we derive approximated imputation models for these scenarios. In particular, only for the latter one, the approximate model is the same as in \cite{beesley2016multiple}. Through simulation studies, we investigate how these two multiple imputation approaches perform in practice in this more general setting, an aspect not explored in the original work by \cite{beesley2016multiple}. The simulation results also emphasise the need to differentiate between the covariates modelling the probability of cure and the survival of the uncured patients, as including all covariates in both components leads to wider confidence intervals. However, model selection is not a trivial task in the presence of missing values and model misspecification seems to deteriorate the performance of the exact imputation method. However, the approximate imputation model appears to be quite robust to model misspecification and would be the recommended choice despite the slightly wider confidence intervals compared to the exact approach. Furthermore, to facilitate the application of our method in practice, we provide general-purpose software\footnote{The source code is available at \url{https://github.com/martacip/mi_curemodels}}.

In Section \ref{section:basic}, we introduce the fundamental concepts underlying cure models and the use of multiple imputation in this context. In Section \ref{section:method}, we outline the methodology for performing multiple imputation in cure models when the sets of covariates for incidence and latency models are different. In Section \ref{section:simulation}, we evaluate the performance of this method using simulated data. In Section \ref{section:data}, we apply the developed methodology to analyse osteosarcoma data from the BO06 clinical trial and compare the results with the complete case approach for sensitivity. Finally, in Section \ref{section:discussion}, we conclude with a discussion.


\section{Basic concepts}
\label{section:basic}

\subsection{Mixture cure models}
\label{subsection:cure}

We are interested in studying the time $T$ until an event of interest occurs, in a sample consisting of cured and uncured individuals. The cured individuals will not experience the event of interest during their lifetime. We denote by $G$ the binary indicator of the uncured status, i.e. $G=1$ if uncured and $G=0$ if cured. However, due to the censoring at time $C$ and to the limited time window covered by the study at hand, these variables cannot be directly observed. Instead, we observe the follow-up time $Y=\min(T, C)$ and the event indicator $\Delta=I(T<C)$. For uncensored observations ($\Delta = 1$), $G = 1$, while for censored observations $G$ is unknown.

We consider a mixture cure model, which expresses overall survival as a combination of the probability of being cured (with potentially infinite survival) and the survival of the uncured individuals. The probability of being uncured, also known as incidence, given a design vector $X \in \mathbb{R}^p$, is denoted by $\pi(X)=P(G=1 \vert X)$, and is typically described by a logistic regression function \citep{legrand2019cure,farewell1977model}
\begin{align}
\label{eqn:pi}
\pi(X)=\frac{\exp{(\alpha_0+\alpha^TX)}}{1+\exp{(\alpha_0+\alpha^TX)}}
\end{align}
where $\alpha_0 \in \mathbb{R}$ is the intercept and $\alpha \in \mathbb{R}^p$ denotes the vector of regression parameters. \\
The survival of the uncured patients, known as latency, conditional on a set of covariates $Z \in \mathbb{R}^q$, which may differ from $X$, is represented by $S_u(t \vert Z)=P(T>t \vert Z, G=1)$, and it can be modelled employing a regression model such as the Cox PH model \citep{cox}
\begin{align}
\label{eqn:S_u}
S_u(t \vert Z)=\exp\left(-H_0(t)\exp{(\beta^T Z)}\right)
\end{align}
where $H_0$ denotes the cumulative baseline hazard function of the uncured and $\beta \in \mathbb{R}^q$ is the vector of regression parameters.
In this model, the proportional hazards assumption holds for the subpopulation of uncured patients, but not for the entire population. Combining these components \citep{kuk1992mixture}, the overall survival is given by
\begin{align}
S(t \vert X,Z)=(1-\pi(X))+\pi(X) S_u(t \vert Z)
\end{align}
This model is referred to as the Cox PH cure model.

For a cure model to be identifiable, a sufficiently long follow-up is required, i.e. the study duration should be longer than the support of the distribution for the event times of the uncured individuals. For a more extensive discussion of the assumption of sufficient follow-up, we refer the reader to the review by \cite{amico2018cure}. In practice, to check whether such an assumption is satisfied, one typically combines medical knowledge with a visual inspection of the Kaplan-Meier estimate. Ideally, we would like to observe a long plateau after the last observed event, which contains many censored observations. Recently, some statistical tests have also been proposed for the sufficient follow-up assumption \citep{assump1, assump2,followup}. 

The complete-data likelihood function for the Cox PH cure model can be written as 
\begin{align}
\label{eqn:like}
\mathcal{L}_{c}(\alpha_0, \alpha, \beta, H_0) = \prod_{i=1}^{n} \left[ \pi(X_i) f_u(Y_i \vert Z_i) \right]^{\Delta_i G_i} \left[ \pi(X_i) S_u(Y_i \vert Z_i) \right]^{(1-\Delta_i) G_i} \left[ 1 - \pi(X_i) \right]^{(1-\Delta_i) (1-G_i)}
\end{align}
and it can be factorized into two components, containing the parameters of the incidence and the latency, respectively. \\
Since this model depends on the latent cure status $G$, the set of parameters is usually estimated using an Expectation-Maximization (EM) algorithm \citep{peng2000nonparametric,sy2000estimation}, which is an iterative two-step procedure.
In the E-step, the posterior expectation of the complete data log-likelihood over the latent uncured indicator is obtained, conditional on the parameters estimated from the previous iteration. For individual $i \in (1,...n)$ at iteration $r$, this is given by
\begin{equation*}
\label{eqn:q_update}
\begin{split}
q_i^{(r)} & = E(G_i \vert Y_i, \Delta_i, X_i, Z_i, 
\alpha_0^{(r-1)}, \alpha^{(r-1)}, \beta^{(r-1)}, H_0^{(r-1)}) \\ 
& = \Delta_i + (1 - \Delta_i) \frac{\pi^{(r-1)}(X_i) S_u^{(r-1)}(Y_i \vert Z_i)} 
{1 - \pi^{(r-1)}(X_i) + \pi^{(r-1)}(X_i) S_u^{(r-1)}(Y_i \vert Z_i)}
\end{split}
\end{equation*}
where $\pi^{(r-1)}$ and $S_u^{(r-1)}$ are computed according to equations \eqref{eqn:pi} and \eqref{eqn:S_u} by using the parameter estimates resulting from iteration $r-1$, namely $\alpha_0^{(r-1)}, \alpha^{(r-1)}, \beta^{(r-1)}, H_0^{(r-1)}$. 
The M-step of the EM algorithm consists of maximising the expected log-likelihood for complete data given in \eqref{eqn:like}, where $G_i$ is replaced by its expectation $q_i^{(r)}$ at each iteration. This reduces to a simpler weighted data problem, with an expected log-likelihood which can be maximised by standard algorithms. \\
If parameter distinctiveness holds, the two components of the expected likelihood can be maximised separately, one with respect to $\alpha_0$ and $\alpha$, and the other with respect to $\beta$ and $H_0$. Maximisation with respect to $H_0$ at iteration $r$ leads to the following modified Breslow estimator \citep{breslow} of the baseline cumulative hazard function 
\begin{align}
\label{eqn:H0}
\hat{H}_0^{(r)}(t)={\sum\limits_{t_j \leq t}} \frac {D(t_j)} {{\sum\limits_{i \in R_j}} q_i^{(r)} \exp{(\beta^{(r-1)T}Z_i)}}
\end{align}
where $D(t_j)$ is the number of events at time $t_j$ and $R_j$ represents the set of observations at risk at time $t_j$ \citep{sy2000estimation}.

\subsection{Handling missing values by multiple imputation}
\label{subsection:missing}

We first illustrate the multiple imputation approach for a general model with outcome $Y$ and design vector $X \in \mathbb{R}^p$, where $X^{(j)}$ has missing observations.
The procedure consists of imputing the incomplete covariate $X^{(j)}$ by defining a model for this element, conditional on the outcome $Y$ and the remaining covariates $X^{(-j)} = X \setminus X^{(j)}$.
This is done using the fully conditional specification (FCS) approach \citep{van2006fully}, which requires the specification of the conditional distribution $f(X^{(j)} \vert Y, X^{(-j)}; \theta$), where $\theta$ represents a set of parameters. Such parameters depend on $j$, but we will omit it from the notation for simplicity. The procedure begins by fitting a regression model for $X^{(j)}$ with covariates $Y$ and $X^{(-j)}$ on the portion of observed data for $X^{(j)}$. This provides us with estimates $\hat{\theta}$ for the regression coefficients $\theta \in \mathbb{R}^p$, as well as the estimated covariance matrix $\hat{\Sigma}_{\theta}$.
We then draw values $\theta^*$ from the posterior distribution of $\theta$ given the observations, which is approximated by a Gaussian distribution, $N(\hat{\theta}, \hat{\Sigma}_\theta)$.
The imputed values for the cases where $X^{(j)}$ is missing are drawn from the conditional distribution $f(X^{(j)} \vert Y, X^{(-j)}; \hat{\theta}^*)$. This step is repeated $K$ times, resulting in $K$ imputed samples.

\begin{figure*}[h]
    \centering
    \includegraphics[height=0.4\textheight]{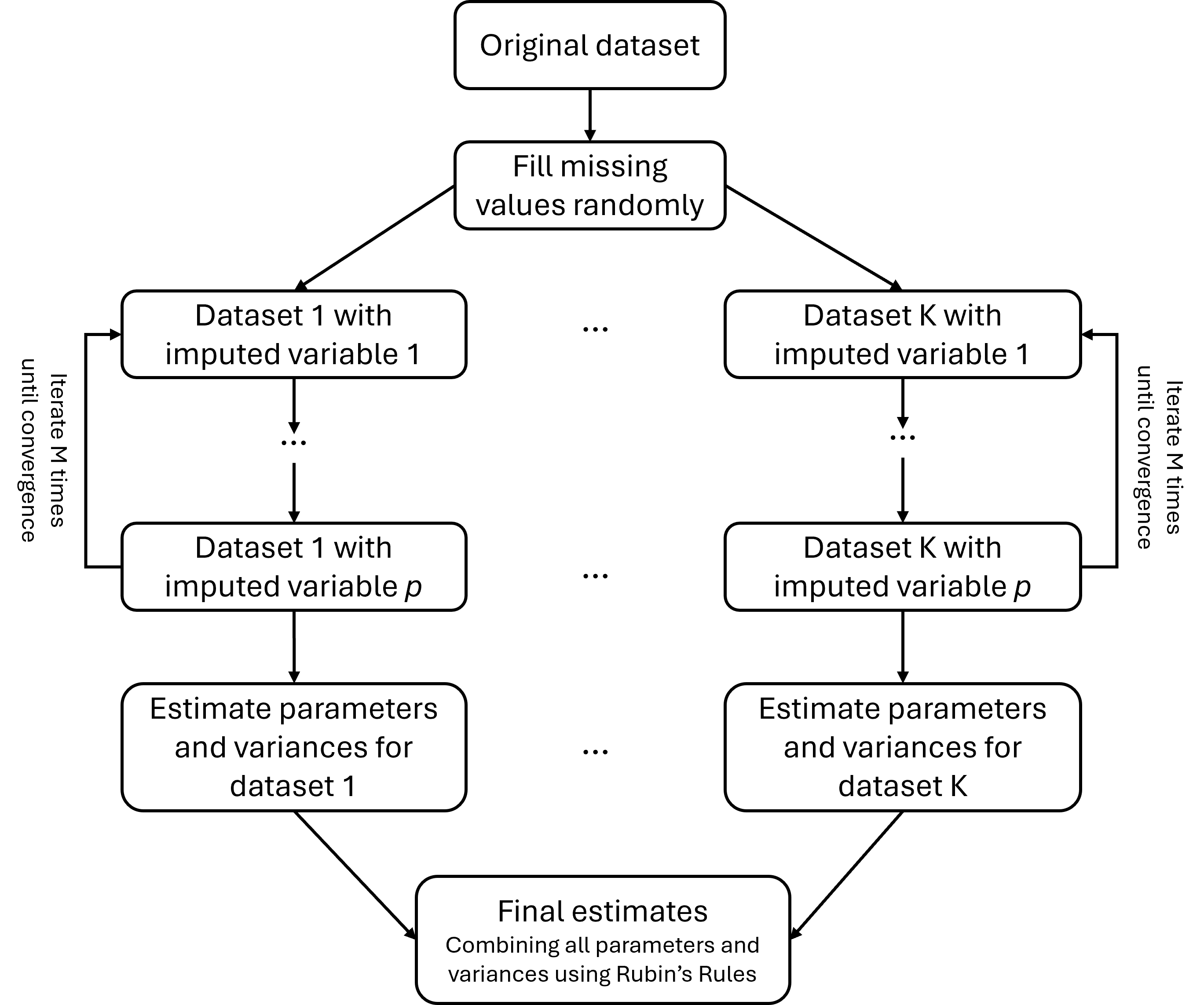}
    \caption{Multiple imputation by chained equations procedure}
    \label{fig:imputation}
\end{figure*}

When there are multiple variables with missing observations, this process is repeated sequentially for each one of them, using the observed and imputed values of the remaining variables.
This procedure, known as multiple imputation by chained equations, is summarised in Figure \ref{fig:imputation}. It starts by filling all missing values randomly and then creates $K$ datasets with imputed values for incomplete variables. The order in which the variables are imputed is determined by the proportion of missing cases, starting with the variable that has the least missing data \citep{van2018flexible}. Each variable uses the observed and most recently imputed values of the remaining ones, and this method is iterated on each dataset until convergence.

After obtaining $K$ imputed datasets, the set of parameters of interest $\Psi$ is estimated separately on each of these datasets, resulting in a vector of estimates $(\hat{\Psi}_1, ..., \hat{\Psi}_K)$ and estimated covariance matrices $(\hat{\Sigma}_{\Psi 1}, ..., \hat{\Sigma}_{\Psi K})$. 
Finally, the pooled estimate is obtained by combining the $K$ estimates using Rubin's rule \citep{rubin2004multiple}
\begin{align*}
\hat{\Psi} = \frac{1}{K} \sum\limits_{k=1}^{K}\Psi_k
\end{align*}
and the final covariance matrix is given by 
\begin{align*}
\hat{\Sigma}_\Psi & = \frac{1}{K} \sum\limits_{k=1}^{K}\Sigma_{\Psi k} + \\
& + \left( 1+\frac{1}{K} \right) \frac{1}{K-1} \sum\limits_{k=1}^{K}(\Psi_k-\hat{\Psi})(\Psi_k-\hat{\Psi})^T
\end{align*}
where the first component is the average of the repeated complete-data posterior covariances of $\Psi$ (within component), and the second component is the estimate of the covariance between the $K$ complete-data posterior means of $\Psi$ (between component).

\subsection{Multiple imputation for cure models}
\label{subsection:multiple}

To perform multiple imputation in the case of a cure model with missing covariates, the first step is to determine the conditional distribution of a covariate with missing values given the outcome and the other covariates. For survival data, the outcome consists of both the follow-up time $Y$ and the event indicator $\Delta$. In the case of the standard Cox model (without cure fraction), it has been shown that a suitable model for imputing the missing covariate values is based on $\Delta$, the cumulative baseline hazard $H_0(Y)$ and the other covariates \citep{white2009imputing}. 

However, for the cure model, an additional challenge arises from the fact that even the cure status $G$ is unobserved for censored observations. For a Cox PH cure model where the design vector is common to both the incidence and the latency (i.e. $X=Z$), \cite{beesley2016multiple} derived the exact conditional distribution based on the complete likelihood function in \eqref{eqn:like}. They proposed a chained equations procedure which imputes iteratively the missing values for the cure status $G$ and the incomplete covariates. 
In practice, drawing observations from the exact conditional distribution, which has a complicated form, is often computationally expensive. To reduce the computational time, \cite{beesley2016multiple} proposed approximations of the exact conditional distribution which are easier to handle in practice.
However, assuming that $X$ and $Z$ are the same is quite restrictive, preventing us from fully exploiting the advantage of a mixture cure model in distinguishing between covariates with a curative or just a life-prolonging effect. 

The ideas of \cite{beesley2016multiple} have been explicitly discussed in this limited context, but we think they can be extended to the more general model described in Section \ref{subsection:cure}, considering three possible scenarios: the incomplete covariate belongs only to $X$, only to $Z$, or to both $X$ and $Z$. The scenario where the variable with missing values belongs to both $X$ and $Z$ is treated in Section \ref{section:method}, while the cases where it belongs to only one of the two components (either the incidence or the latency one) are discussed in the appendices. Specifically, Appendix A deals with the case using the exact conditional distribution, while Appendix B addresses the case of the approximate conditional distribution. Further details about the method proposed by \cite{beesley2016multiple} and how it relates to this work are given in Section \ref{section:method}.

\section{Methodology}
\label{section:method}

In this section, we describe the multiple imputation procedure in cure models with incomplete covariates. As mentioned in Section \ref{subsection:missing}, the main aspect of the multiple imputation procedure involves generating imputed values based on the distribution of the incomplete variable conditional on the other ones. Therefore, we first present the exact form of the conditional distribution and then outline the steps of the algorithm. Finally, we introduce an approximation of the conditional distribution that enables us to generate imputed values at a reduced computational cost.

\subsection{Using the exact conditional distribution}
\label{subsection:exact}

Let $W \in \mathbb{R}$ be the incomplete variable present in both the incidence and the latency model. The results for the cases where $W$ belongs either to the incidence or to the latency model are discussed in the Appendix. 
Since $W$ belongs to both $X$ and $Z$, there exist $1\leq j\leq p$ and $1\leq l\leq d$ such that $W=X^{(j)}=Z^{(l)}$. We denote the set of incidence and latency covariates excluding $W$ as $X^{(-j)} = X \setminus X^{(j)}$ and $Z^{(-l)} = Z \setminus Z^{(l)}$ respectively.
To detail the proposed procedure in a more general context, we will consider two different choices for $W$, either having a Gaussian or a Bernoulli distribution, conditional on the other covariates. However, results can be similarly generalized to other distributions. We note also that the following formulas can be obtained from equation (2) in \cite{beesley2016multiple} by including all covariates to both parts of the model with corresponding coefficients equal to zero for the non-active covariates. However, in terms of estimation process, it makes a difference whether a certain covariate is included or not in each submodel.

\textit{Case 1.} Let us first assume that $W \vert X^{(-j)}, Z^{(-l)}$ follows a normal distribution, and that, at least approximately, the dependence of $W$ on $X^{(-j)}$ and $Z^{(-l)}$ is summarized by a change in the location (mean) parameter only. That is, we specify: 
\begin{align*}
    W \vert X^{(-j)}, Z^{(-l)} \sim \text{N}(\mu,\sigma),
\end{align*} 
where
\begin{align}
\label{eqn:mu}
\mu = \theta_0 + \sum\limits_{b \neq j}^{p} \theta_b X_b + \sum\limits_{\substack {s \neq l \\ Z_s \notin X}}^{d}{\theta_{p+s}} Z_s
\end{align}
Note that we only consider variables in $Z$ that are not in $X$ ($Z_s \notin X$) to prevent any useless overlapping information.
In this case, the complete data likelihood is given by
\begin{align}
\label{eqn:L_Norm}
\mathcal{L}(\alpha_0, \alpha, \beta, H_0, \mu, \sigma) 
& \, = \prod_{i=1}^{n}
\left[\pi(X_i) f_u(Y_i \vert Z_i) \right] ^ {\Delta_i G_i}
\left[\pi(X_i) S_u(Y_i \vert Z_i) \right] ^ {(1-\Delta_i) G_i} \notag \\ 
& \, \times \left[1-\pi(X_i) \right] ^ {(1-\Delta_i) (1-G_i)} 
f(W_i \vert X_{i}^{(-j)}, Z_{i}^{(-l)}) \notag \\
& \, =\prod_{i=1}^{n}
\left[\frac{e^{\alpha_0+\alpha^TX_i}}{1+e^{\alpha_0+\alpha^TX_i}}
\left( h_0(Y_i) e^{\beta^T Z_i} \right)^{\Delta_i} e^{-H_0(Y_i) e^{\beta^T Z_i}}
\right]^{G_i} \notag \\
& \, \times \left(\frac{1}{1+e^{\alpha_0+\alpha^TX_i}}
\right)^{1-G_i}
\left( e^\frac{-(W_i-\mu_i)^2}{2\sigma^2} \right)
\end{align}
Therefore, the exact distribution of $W$ conditional on all the remaining covariates is as follows
\begin{align}
\label{eqn:fW_Norm}
 f(W_i \vert X_{i}^{(-j)}, Z_{i}^{(-l)}, Y_i, \Delta_i, G_i) 
 \propto \left( \frac{e^{\alpha_0+\alpha^TX_i}}{1+e^{\alpha_0+\alpha^TX_i}}
e^{\beta^T Z_i \Delta_i}
e^{-H_0(Y_i) e^{\beta^T Z_i}}
\right)^{G_i} \left( \frac{1}{1+e^{\alpha_0+\alpha^TX_i}}
\right)^{1-G_i}
\left( e^\frac{-(W_i-\mu_i)^2}{2\sigma^2} \right)
\end{align}
One can draw values to impute $W$ from this distribution using a Metropolis-Hasting algorithm \citep{metropolis1953equation,hastings1970monte}. This algorithm can draw values from a given generic probability distribution (the target distribution) by sampling from a Markov chain whose stationary distribution is the target distribution. \\

\textit{Case 2.} Alternatively, $W \vert X^{(-j)}, Z^{(-l)}$ could be a categorical random variable. For instance, if $W$ is binary, we can specify:
\begin{align*}
    W \vert X^{(-j)}, Z^{(-l)} \sim \text{Ber}(\expit(\mu)),
\end{align*}
where $\expit(x)=e^x/(1+e^x)$ and $\mu$ is given by \eqref{eqn:mu}. The complete data likelihood is as in \eqref{eqn:L_Norm} where $f(W_i \vert X_{i}^{(-j)}, Z_{i}^{(-l)})$ is replaced by $P(W_i=w_i \vert X_{i}^{(-j)}, Z_{i}^{(-l)})$ with $w_i\in\{0,1\}$. Recalling that $\logit(x) = \log\bigl( \frac{x}{1-x}\bigr)$, the exact distribution of $W$ conditional on all the remaining covariates takes the form
\begin{align}
\label{eqn:fW_Bin}
\logit \left[ P(W_i=1 \vert X_{i}^{(-j)}, Z_{i}^{(-l)}, Y_i, \Delta_i, G_i) \right] & = \log \left[ \frac{P(W_i=1 \vert X_{i}^{(-j)}, Z_{i}^{(-l)}, Y_i, \Delta_i, G_i)}{P(W_i=0 \vert X_{i}^{(-j)}, Z_{i}^{(-l)}, Y_i, \Delta_i, G_i)} \right] 
\notag \\
& = G_i \Delta_i \beta_l
- G_i H_0(Y_i) e^{\sum_{s \neq l}^{d}{\beta_s Z_{i,s}}} (e^{\beta_l}-1)
+ G_i \alpha_j \notag \\
&  + \log \left( 1+e^{\alpha_0+\sum_{b \neq j}^{p}{\alpha_b X_{i,b}}} \right) - \log \left(1+e^{\alpha_0+\alpha_j+\sum_{b \neq j}^{p}{\alpha_b X_{i,b}}} \right) + \mu_i
\end{align}

Given that the conditional distributions depend on $\mu$ in both cases, one must draw values $\theta^*$ for the parameters $\theta$ before drawing values for $W$. This is done by running a regression model (logistic regression when $W \vert X^{(-j)}, Z^{(-l)} \sim$ Bernoulli and linear regression when $W \vert X^{(-j)}, Z^{(-l)} \sim$ Normal), obtaining point estimates $\hat{\theta}$ and the corresponding covariance matrix estimates $\hat{\Sigma}_\theta$. Then, we draw values $\theta^*$ from a Normal$\,(\hat{\theta}, \hat{\Sigma}_\theta)$ distribution.

\subsection{Implementing multiple imputation in cure models}
\label{subsection:multcure}

Let us consider the Cox PH cure model described in Section \ref{subsection:cure}. To impute missing observations for any incomplete covariate by drawing values from its conditional distribution, we need to know $G$ and all the model parameters. Therefore, we use an iterative procedure where we first estimate the parameters, impute $G$, and then impute the missing covariate values. The pseudo-code of the corresponding algorithm is provided in Algorithm~\ref{alg:cap}, and it can be briefly described as follows:
\begin{algorithm*}[h]
\caption{Imputation of missing covariates using exact conditional distribution}\label{alg:cap}
\begin{algorithmic}
\Require dataset of $n > 0$ observations; number of iterations for chained equations $M > 0$ ; number of imputed datasets $K > 0$; $W \vert X^{(-j)}, Z^{(-l)} \sim N(\mu, \sigma)$ \, or \, $W \vert X^{(-j)}, Z^{(-l)} \sim Ber(\expit(\mu))$
\Ensure Imputed data for W

\State \textbf{0: Initialization ($m=0$)}
\State Assign values to $\alpha_{0}^{(0)}, \alpha^{(0)}, \beta^{(0)}, H_{0}^{(0)}(t)$, and randomly fill the missing cases of $W$ producing $W^{(0)}$

\For{$m=1,\dots,M$}

\State \textbf{1: Estimate $H_{0}(t)$}
\For{$i=1,\dots,n$}
\State \textbf{a:} Compute $\pi^{(m-1)}(X_{i}) = \frac{\exp(\alpha_{0}^{(m-1)} + \alpha_{W}^{(m-1)}W_{i}^{(m-1)} + \alpha_{X}^{(m-1)T}X_{i}^{(-j)})}{1 + \exp(\alpha_{0}^{(m-1)} + \alpha_{W}^{(m-1)}W_{i}^{(m-1)} + \alpha_X^{(m-1)T}X_{i}^{(-j)})}$
\State \textbf{b:} Compute $S_{u}^{(m-1)}(t \vert Z_{i}) = \exp\bigl (-H_{0i}^{(m-1)}(t) \exp(\beta_W^{(m-1)}W_{i}^{(m-1)} + \beta_Z^{(m-1)T} Z_{i}^{(-l)}) \bigr)$
\State \textbf{c:} Compute $q_{i}^{(m)} = \Delta_{i} + (1-\Delta_{i})\frac{\pi^{(m-1)}(X_{i}) S_{u}^{(m-1)}(t \vert Z_{i})}{1-\pi^{(m-1)}(X_{i}) + \pi^{(m-1)}(X_{i}) S_{u}^{(m-1)}(t \vert Z_{i})}$
\EndFor
\State \quad Get $H_{0}^{(m)}(Y_i) = \sum_{t_{j}<t_{i}} \frac{D(t_{j})}{\sum_{i \in R_{j}} q_{i}^{(m)} \exp(\beta_W^{(m-1)}W_{i}^{(m-1)} + \beta_Z^{(m-1)T} Z_{i}^{(-l)})}$

\vspace{0.2cm}

\State \textbf{2: Draw values for $\alpha_{0}, \alpha, \beta$}
\State \quad \textbf{a:} Compute $\hat{\alpha_{0}}, \hat{\alpha}$ and $\hat{\Sigma}_{(\alpha_{0}, \alpha)}$ from $\logit[P(G^{(m-1)}=1 \vert W^{(m-1)}, X^{(-j)})]$
\State \quad \textbf{b:} Draw $(\alpha_{0}^{(m)}, \alpha^{(m)})$ from MVN$\bigl( (\hat{\alpha_{0}}, \hat{\alpha}), \hat{\Sigma}_{(\alpha_{0}, \alpha)}\bigr) $
\State \quad \textbf{c:} Compute $\hat{\beta}$ and $\hat{\Sigma}_{\beta}$ fitting $S_{u}(t \vert W^{(m-1)}, Z^{(-l)})$ to the subjects such that $G_{i}=1$
\State \quad \textbf{d:} Draw $\beta^{(m)}$ from MVN$(\hat{\beta}, \hat{\Sigma}_{\beta})$

\vspace{0.2cm}

\State \textbf{3: Impute $G$}
\ForAll{$G_{i} =$ \textit{NA} } draw from $\logit [P(G_{i} = 1 \vert W_{i}^{(m-1)}, X_{i}^{(-j)}, Z_{i}^{(-l)}, Y_{i}, \Delta_{i}=0; \alpha_{0}^{(m)}, \alpha^{(m)}, \beta^{(m)})]$
\EndFor 

\vspace{0.2cm}

\State \textbf{4: Impute W}
\State Specify $f(W_{i} \vert G_{i}, \Delta_{i}, Y_{i}, X_{i}^{(-j)}, Z_{i}^{(-l)}; \theta)$ and
\If{$W \sim Ber(\expit(\mu))$}
    \State \textbf{a:} Compute $\hat{\theta}$ and $\hat{\Sigma}_{\theta}$ from a logistic regression on $W_{i}^{(m-1)}$ using $X_{i}^{(-j)}, Z_{i}^{(-l)}$
    \State \textbf{b:} Draw values $\theta^*$ from MVN$(\hat{\theta}, \hat{\Sigma}_{\theta})$
    \State \textbf{c:} \textbf{for all} $W_{i}^{(0)} =$ \textit{NA} \textbf{do} draw imputed values from $\logit[ P(W_{i} = 1 \vert G_{i}, \Delta_{i}, Y_{i}, X_{i}^{(-j)}, Z_{i}^{(-l)}; \theta^*)]$
\Else \If{$W \sim N(\mu, \sigma)$}
    \State \textbf{a:} Compute $\hat{\theta}$ and $\hat{\Sigma}_{\theta}$ from a linear regression on $W_{i}^{(m-1)}$ using $X_{i}^{(-j)}, Z_{i}^{(-l)}$
    \State \textbf{b:} Draw values $\theta^*$ from MVN$(\hat{\theta}, \hat{\Sigma}_{\theta})$
    \State \textbf{c:} \textbf{for all} $W_{i}^{(0)} =$ \textit{NA} \textbf{do} Metropolis-Hastings draw using a random normal walk centered at $W_{i}^{(m-1)}$
\EndIf
\EndIf

\If{$W \in \mathbb{R}^{a}$ with $a > 1$} \State repeat \textbf{a}, \textbf{b} and \textbf{c} for every $W$
\EndIf

\EndFor

\vspace{0.2cm}

\State \textbf{Repeat} steps \textbf{1}, \textbf{2}, \textbf{3} and \textbf{4} to obtain $K$ complete datasets 

\end{algorithmic}
\end{algorithm*}

\begin{enumerate}
\item[0.] The algorithm is initialized with values for $\alpha_0$, $\alpha$ and $\beta$ obtained by applying a Cox PH cure model to the complete-case dataset. By using a Breslow estimator \citep{breslow} on the same data, we obtain the initial values for $H_{0}(t)$. The missing cases of $W$ are randomly filled in. 
\item The first step consists in estimating $H_0(Y)$, the baseline cumulative hazard rate for the individual times $(Y_1,...,Y_n)$ using the estimator in \eqref{eqn:H0}.
\item In the second step, we compute the estimates of the model parameters $\alpha_{0}$, $\alpha$ and $\beta$. For $\alpha_0$ and $\alpha$, this is done by fitting a logistic model with outcome $G$ and covariates $X$. Initially, the cure status $G$ has missing values for those censored units not in the right tail of the survival distribution (see step 3). In subsequent iterations, $G$ is filled in with the imputations obtained at the previous iteration. We then draw values $(\alpha_0^*,\alpha^*)$ from a multivariate Gaussian distribution, by using the asymptotic distribution of the ML estimates, e.g. MVN$\,((\hat{\alpha}_0,\hat{\alpha}), \hat{\Sigma}_{(\alpha_0,\alpha)})$. For $\beta$, this is done by fitting a Cox model to the observations where $G=1$, with follow-up times $Y$, event indicator $\Delta$, and covariates $Z$. We draw values $\beta^*$ from MVN$\,(\hat{\beta}, \hat{\Sigma}_\beta)$.
\item The following step consists of imputing the cure status $G$, which is known to be 1 for uncensored observations and assumed to be 0 for censored individuals with follow-up times after a certain cut-off point. As suggested in the literature \citep{sy2000estimation}, we take the cut-off point equal to the last uncensored follow-up time. For censored observations until the cut-off point, $G$ is assumed to be missing at random conditional on $Y$ and $\Delta$ \citep{beesley2016multiple}, and it can be imputed using its exact conditional distribution
\begin{align*}
\logit \left[ P(G_i=1 \vert X_i, Z_i, Y_i, \Delta_i=0) \right] = \log \left[ \frac{S_u(Y_i \vert Z_i) \pi(X_i)}{1-\pi(X_i)}  \right] = - H_0(Y_i) 	 e^{\beta^{T}Z_i} + \left( \alpha_0+\alpha^{T}X_{i} \right)
\end{align*}
which is derived from the likelihood introduced in \eqref{eqn:like}. For the sake of clarity, let us recall that both $Z$ and $X$ include the variable with missing cases $W$.
\item Finally, we impute each missing covariate sequentially using the conditional distributions given in \eqref{eqn:fW_Norm} or \eqref{eqn:fW_Bin}.
\end{enumerate}
Each step makes use of the most recently imputed values of the incomplete variables and the most recently estimated model parameters.

\subsection{Approximation of the conditional distribution}

Drawing values from the exact conditional distributions in \eqref{eqn:fW_Norm} and \eqref{eqn:fW_Bin} can be computationally intensive. Indeed, M-H algorithms
are slowed down by the computation of complex target distributions and due to the slow convergence of such algorithms, one may need a very large number of iterations. \cite{beesley2016multiple} proposed an approximation and compared it to three alternatives from the literature. A simulation study indicated that the proposed approximation outperforms the others in terms of bias when the data has a cure fraction and $X=Z$. We extend this approach to cases where the sets of incidence and latency covariates differ. As in \cite{beesley2016multiple}, we make use of the first-order expansion
\begin{align}
\label{eqn:approx1}
\log(1+z) \approx \log(1+c) + \frac{z-c}{1+c}
\end{align}
if $z$ is near $c$, and
\begin{align}
\label{eqn:approx2}
e^{a X+b Y} \approx e^{a\bar{X}+b\bar{Y}} [1+a(X-\bar{X}) + b(Y-\bar{Y})]
\end{align}
if $Var(a X+b Y)$ is small. The only constraint to consider for the approximation \ref{eqn:approx1} to hold and be valid is that the argument of the logarithm function, $1+z$ and $1+c$, must be positive. Therefore, $z > -1$ and $c > -1$. \\
First, we start discussing the case when $W \vert X^{(-j)}, Z^{(-l)} \sim N(\mu,\sigma)$. The $\log$ of the exact conditional distribution of $W$ is given by
\begin{align*}
\log \left( f(W_i \vert X_{i}^{(-j)}, Z_{i}^{(-l)}, Y_i, \Delta_i, G_i) \right) & = G_i \Delta_i \beta_l W_i
- G_i H_0(Y_i) e^{\beta^T Z_i}
+ G_i \alpha_j W_i \notag \\
& - \log \left( 1 + e^{\alpha_0+\alpha^T X_i} \right)
- \frac{1}{2 \sigma^2} \left( W_i - \mu_i \right)^2
+C
\end{align*}
where $C$ is a general additive term constant on $W_i$. Applying \eqref{eqn:approx1}, \eqref{eqn:approx2}, and a second-order Taylor approximation, one obtains
\begin{align*}
\log \left( f(W_i \vert X_{i}^{(-j)}, Z_{i}^{(-l)}, Y_i, \Delta_i, G_i) \right)
& \approx - \frac{1}{2 \sigma^2} W_i^2 + \left[ G_i \Delta_i \beta_l
- G_i H_0(Y_i) e^{\beta^T \bar{Z}} \beta_l\right] W_i
\tilde{C} \notag \\
& + \left[ G_i \alpha_j 
- \frac{e^{\alpha_0 + \alpha^T \bar{X}}}{1+e^{\alpha_0 + \alpha^T \bar{X}}} \alpha_j
+ \frac{1}{\sigma^2} \mu_i
\right] W_i
\tilde{C}
\end{align*}
where $\tilde{C}$ is another constant with respect to $W_i$.
Recalling that $\mu_i$ is a linear combination of $X_{i}^{(-j)}$ and $Z_{i}^{(-l)}$, one can see that the distribution of $W_i \vert X_{i}^{(-j)}, Z_{i}^{(-l)}, Y_i, \Delta_i, G_i$ is approximately Normal, where the mean is a linear combination of $X_{i}^{(-j)}$, $Z_{i}^{(-l)}$, $G_i$, $G_i \Delta_i$ and $G_i H_0(Y_i)$. Then, instead of using the known expressions for the coefficients of this approximate linear approximation, we fit a linear regression model with these covariates to approximate the exact distribution of $W$.

Secondly, if we assume that $W \vert X^{(-j)}, Z^{(-l)} \sim Ber(\expit(\mu))$, the conditional distribution of $W_i \vert X_{i}^{(-j)}, Z_{i}^{(-l)}, Y_i, \Delta_i, G_i$ can be approximated using \eqref{eqn:approx1} and \eqref{eqn:approx2}, obtaining
\begin{align*}
\logit \left[ P(W_i=1 \vert X_{i}^{(-j)}, Z_{i}^{(-l)}, Y_i, \Delta_i, G_i) \right]
& \approx G_i \Delta_i \beta_l - G_i H_0(Y_i) (e^{\beta_l}-1) e^{\sum_{s \neq l}^{d}{\beta_s \bar{Z}_{s}}}
\left[ 1 + \sum_{s \neq l}^{d}{\beta_s \left( Z_{i,s} - \bar{Z}_{s} \right)} \right] \notag \\
& + G_i \alpha_j
+ \frac{e^{\alpha_0+\sum_{b \neq j}^{p}{\alpha_b \bar{X}_b}}}{1+e^{\alpha_0+\sum_{b \neq j}^{p}{\alpha_b \bar{X}_b}}} 
\left[ 1 + \sum_{b \neq j}^{p}{\alpha_b \left( X_{i,b} - \bar{X}_{b} \right)} \right] \notag \\
& - \frac{e^{\alpha_0+\alpha_p+\sum_{b \neq j}^{p}{\alpha_b \bar{X}_b}}}{1+e^{\alpha_0+\alpha_p+\sum_{b \neq j}^{p}{\alpha_b \bar{X}_b}}} 
\left[ 1 + \sum_{b \neq j}^{p}{\alpha_b \left( X_{i,b} - \bar{X}_{b} \right)} \right] + \mu_i + C
\end{align*}
This is a linear combination of $X_{i}^{(-j)}$, $Z_{i}^{(-l)}$, $G_i$, $G_i \Delta_i$, $G_i H_0(Y_i)$ and $G_i H_0(Y_i) Z_{i}^{(-l)}$. Therefore, the exact distribution of $W$ can be approximated by fitting a logistic regression model with these covariates as the last step of the imputation procedure presented in Section \ref{subsection:multcure}. \\
Details about the scenarios where the variable with missing values $W$ belongs only to the incidence or the latency model are provided in Appendix  B. If $W$ belongs only to the incidence component, the conditional distribution of $W$ can be approximated by linear or logistic regression with covariates $X_{i}^{(-j)}$, $Z_{i}^{(-l)}$, and $G_{i}$. When $W$ belongs only to the latency, the conditional distribution of W can be approximated by linear or logistic regression with covariates $X_{i}^{(-j)}$, $Z_{i}^{(-l)}$, $G_{i}\Delta_{i}$, $G_{i}H_0(Y_{i})$, and (for \textit{Case 2}) $G_{i}H_{0}(Y_{i})Z_{i}^{(-l)}$.

\subsection{Variable selection in the presence of missing data}
Variable selection in cure models has been addressed in the literature, with approaches such as the penalized likelihood methods proposed by \cite{liu} and \cite{masud}. However, these methods do not account for missing data in covariates, leaving a critical gap in the current methodologies. Addressing this issue accurately is a complex challenge, which goes beyond the scope of this work. 
Here, we aim to provide preliminary suggestions by integrating a regularized Cox cure model \citep{masud} with guidelines from the literature on variable selection in the presence of missing data \citep{zhao}. 

Statistical methods for variable selection combined with imputation generally fall into three primary strategies:
\begin{itemize}
    \item \textbf{Impute-and-then-select:} Apply a variable selection method (e.g., Lasso) to each imputed dataset separately and aggregate the selected variables across all imputed datasets.
    \item \textbf{Stacked imputation:} Stack all imputed datasets into one combined dataset and perform variable selection as if the stacked data were a single dataset. This approach, while computationally efficient, may overstate the precision of variable selection by treating the stacked data as independent.
    \item \textbf{Bootstrap-and-impute:} Combine bootstrap resampling with imputation to account for both the variability in the imputed datasets and the sampling variability in the variable selection process.
\end{itemize}

A commonly used method in the standard settings is the one proposed by \cite{wood}, which performs model selection on each imputed dataset separately. Hence, in the context of a mixture cure model, we propose to incorporate a regularized Cox cure model \citep{masud} into the multiple imputation process. Specifically, the variable selection process is repeated at each imputation step, using a penalized likelihood approach. The final set of selected variables is then defined as those selected in at least $\gamma M$ of the $M$ imputed datasets, where $\gamma$ is a user-defined threshold (e.g., $\gamma = 0.5$).

Nonetheless, significant challenges remain. One major hurdle is the determination of tuning parameters, such as the penalization coefficient in Lasso, which typically requires cross-validation. Apart from being computationally intensive, cross-validation introduces additional variability and complicates the integration with multiple imputation. Moreover, the issue of post-selection inference - ensuring valid statistical inference after the variable selection process - remains unresolved in this context and is rarely addressed in existing studies on cure models. 
While this proposal provides a starting point for combining variable selection and imputation in cure models, a comprehensive investigation of these challenges is beyond the scope of this article. Our discussion here serves as an initial framework for future studies to build upon.


\section{Simulation}
\label{section:simulation}
\subsection{Simulation design}
\label{subsection:simul}

We conducted a simulation study to compare the performance of the imputation methods for cure models presented in Section \ref{section:method}. The comparison was carried out considering several scenarios, defined by varying the missing data percentages, model parameters, and variables' distributions.

In this simulation study, we generated 1000 datasets, each containing 500 observations.
The variables used in the study are $W$, $X$ and $Z$. Both $X$ and $Z$ follow a Bernoulli distribution with a probability of 0.5, while $W$ is either $Ber(0.5)$ or $N(0.5,1)$. The choice of a Normal variable with a mean of 0.5 rather than a standard Normal was made to ensure the stability of the cure rate when varying W's distribution. The three variables have a correlation coefficient of 0.5. \\
Variable $X$ is a covariate in the incidence model, $Z$ is a covariate in the latency model, and both have no missing values. Variable $W$ is a covariate in the models for both incidence and latency, and it is either missing completely at random (MCAR) with 15\% or 30\% missing values or missing at random (MAR) with 30\% missing observations. To produce a MAR mechanism, we used a multivariate amputation procedure \citep{amputation} in which the probability of $W$ to be missing depends on both $X$ and $Z$ and the outcome variable ($Y,\Delta)$.

The time to event $T$ is generated from a Weibull PH model of the form
\begin{align*}
S(t)=\exp{(-\lambda t^\rho \exp{(\beta_{W}W+\beta_{Z}Z)})}
\end{align*}
with $\lambda=0.25$, $\rho=1.45$, truncated at $t=8$, to satisfy the sufficient follow-up assumption. For cured individuals (having $G=0$), $T$ is set to an extremely large value, as they never experience the event of interest. \\
The time to censoring $C$ is drawn from an exponential distribution with a parameter equal to 0.08 when $W$ is Bernoulli distributed, and to 0.1 when $W$ is Normally distributed. Both censoring distributions are truncated at 10. \\
The complete uncured indicator $G$ is generated as a Bernoulli random variable with probability $\pi$ defined by a logistic regression 
\begin{align}
\label{G}
P(G=1)=\frac{\exp{(\alpha_0+\alpha_{W}W+\alpha_{X}X)}}{1+\exp{(\alpha_0+\alpha_{W}W+\alpha_{X}X)}}
\end{align}
In practice, we only use the complete indicator drawn from \eqref{G} for generating the follow-up times $Y$, namely to assign large values of $Y$ to cured individuals. Instead, when performing multiple imputation in the cure model, we use the observed uncured indicator, which is 1 for uncensored observations, 0 for observations in the plateau, and unknown otherwise.

We consider two sets of parameter values for the incidence and latency models.
The first one is $\alpha_0=1$, $\alpha_{W}=-1$, $\alpha_X = 0.5$, $\beta_W=-0.2$ and $\beta_Z=0$. This set of parameters is based on the motivating application and therefore is only used when $W$ is Bernoulli distributed with around 15\% of observations missing completely at random.
This scenario leads to an average cure rate of $\sim 33$\%, an average censoring rate of $\sim 45$\%, and an average fraction of observations in the plateau of $\sim 18$\%. \\
The second set of parameters is $\alpha_0=0.1$ and $\alpha_W=\alpha_X=\beta_W=\beta_Z=0.5$. These values are chosen to illustrate a more general setting. For this choice of parameters, we consider $W$ as Normally distributed with around 30\% of observations missing at random or missing completely at random.
Both the scenarios in which the missing mechanism is MCAR and MAR lead to an average cure rate of $\sim 36$\%, an average censoring rate of $\sim 46$\%, and an average fraction of observations in the plateau of $\sim 17$\%.

To summarize this simulation design, the three considered scenarios are the following: 
\begin{itemize}
    \item \textbf{Scenario A:} MCAR data with 15\% missingness; \\ $W \sim$ Bernoulli; $\alpha_0=1$, $\alpha_W=-1$, $\alpha_X=0.5$,  $\beta_W=-0.2$ and $\beta_Z=0$.
    \item \textbf{Scenario B:} MCAR data with 30\% missingness; \\ $W \sim$ Normal; $\alpha_0=0.1$ and $\alpha_W=\alpha_X=\beta_W=\beta_Z=0.5$.
    \item \textbf{Scenario C:} MAR data with 30\% missingness;\\ $W \sim$ Normal; $\alpha_0=0.1$ and $\alpha_W=\alpha_X=\beta_W=\beta_Z=0.5$.
\end{itemize}

For the multiple imputation procedure using the exact conditional distribution in the case of Gaussian $W$, we use a Normal distribution with unit variance in the Metropolis-Hastings sampling algorithm. The sampler runs for 500 iterations before recording results with 100 iterations between successive samples. \\
For the multiple imputation procedure using the approximate conditional distributions, we use the {\fontfamily{qcr}\selectfont mice} \citep{mice} R package. For both the exact and the approximated method, we set the number of iterations of the chained-equations to 10 \citep{iter} and the number of imputed datasets to 10. \\
There are two packages to estimate cure models in R, {\fontfamily{curephEM}\selectfont smcure} \citep{cai2012smcure} and {\fontfamily{curephEM}\selectfont curephEM} \citep{cureph}. We used the latter with the default values for the convergence criteria and bootstrap sampling for standard error estimation. 


\subsection{Simulation results}
In Table \ref{cens}, we present the results of the simulations for the three different scenarios (A-C). 
\begin{sidewaystable*}[p]
\caption{Simulation study results. Evaluation metrics, $B=1000$ samples} 
\centering
 \label{cens} 
\begin{tabular}{lccccccc}
\hline 
\hline \\[-1.8ex]  
\multicolumn{6}{c}{\quad \quad \quad \quad \quad \quad \textbf{MSE$^{1}$ (CI width$^{2}$) Coverage$^{3}$}} \\
 \textbf{Method} & $\alpha_{0}$ & $\alpha_{W}$ & $\alpha_{X}$ & $\alpha_{Z}$ & $\beta_{W}$ & $\beta_{X}$ & $\beta_{Z}$ \\
 \hline \\[-1.8ex] 
 \textbf{Scenario A} & & & & & \\
 \textit{Full data} & 0.04 (0.80) 0.96 & 0.05 (0.92) 0.95 & 0.05 (0.90) 0.96 & & 0.02 (0.53) 0.95 & & 0.02 (0.51) 0.96 \\
 \textit{Complete-case} & 0.05 (0.87) 0.95 & 0.06 (1.00) 0.96 & 0.06 (0.98) 0.96 & & 0.02 (0.58) 0.95 & & 0.02 (0.56) 0.95 \\
 \textit{Exact} & 0.04 (0.83) 0.96 & 0.06 (1.00) 0.97 & 0.05 (0.91) 0.96 & & 0.02 (0.57) 0.96 & & 0.02 (0.51) 0.96 \\
 \textit{Approximate} & 0.04 (0.84) 0.95 & 0.06 (1.00) 0.96 & 0.05 (0.91) 0.96 & & 0.02 (0.57) 0.95 & & 0.02 (0.51) 0.96 \\
 & & & & & \\
 \textbf{Scenario B} & & & & & \\
 \textit{Full data} & 0.03 (0.63) 0.94 & 0.01 (0.46) 0.94 & 0.05 (0.86) 0.96 & & 0.01 (0.29) 0.94 & & 0.02 (0.53) 0.95 \\
 \textit{Complete-case} & 0.04 (0.75) 0.95 & 0.02 (0.55) 0.96 & 0.07 (1.04) 0.96 & & 0.01 (0.35) 0.94 & & 0.03 (0.63) 0.94 \\
 \textit{Exact} & 0.03 (0.65) 0.94 & 0.01 (0.60) 0.99 & 0.04 (0.86) 0.97 & & 0.01 (0.35) 0.94 & & 0.02 (0.54) 0.94 \\
 \textit{Approximate} & 0.03 (0.66) 0.95 & 0.02 (0.60) 0.96 & 0.06 (0.89) 0.94 & & 0.01 (0.35) 0.94 & & 0.02 (0.56) 0.95 \\
 & & & & & \\
 \textbf{Scenario C} & & & & & \\
 \textit{Full data} & 0.03 (0.63) 0.95 & 0.01 (0.46) 0.94 & 0.05 (0.86) 0.97 & & 0.01 (0.29) 0.96 & & 0.02 (0.53) 0.94 \\
 \textit{Complete-case} & 0.35 (0.87) 0.32 & 0.03 (0.68) 0.93 & 0.12 (1.28) 0.94 & & 0.01 (0.33) 0.96 & & 0.03 (0.60) 0.93 \\
 \textit{Exact} & 0.03 (0.65) 0.94 & 0.01 (0.60) 0.99 & 0.04 (0.86) 0.97 & & 0.01 (0.34) 0.93 & & 0.02 (0.54) 0.94 \\
 \textit{Approximate} & 0.03 (0.68) 0.95 & 0.03 (0.79) 0.97 & 0.05 (0.91) 0.96 & & 0.01 (0.35) 0.93 & & 0.02 (0.55) 0.95 \\
 & & & & & \\
 \textbf{Scenario D} & & & & & & & \\
 \textit{Full data} & 0.03 (0.68) 0.93 & 0.02 (0.53) 0.95 & 0.06 (0.96) 0.95 & 0.06 (0.97) 0.96 & 0.01 (0.32) 0.96 & 0.03 (0.59) 0.94 & 0.03 (0.59) 0.95 \\
 \textit{Complete-case} & 0.37 (0.96) 0.41 & 0.05 (0.78) 0.93 & 0.16 (1.46) 0.95 & 0.14 (1.44) 0.95 & 0.01 (0.37) 0.94 & 0.03 (0.68) 0.94 & 0.04 (0.69) 0.95 \\
 \textit{Exact} & 0.04 (0.74) 0.94 & 0.02 (0.63) 0.98 & 0.06 (0.91) 0.94 & 0.06 (0.91) 0.94 & 0.01 (0.35) 0.92 & 0.03 (0.56) 0.93 & 0.03 (0.60) 0.94 \\
 \textit{Approximate} & 0.04 (0.79) 0.96 & 0.05 (0.86) 0.97 & 0.06 (1.00) 0.96 & 0.06 (1.01) 0.96 & 0.01 (0.38) 0.95 & 0.03 (0.59) 0.96 & 0.03 (0.59) 0.96 \\
 & & & & & & & \\
 \hline 
\hline \\[-1.8ex] 
\multicolumn{5}{l}{\scriptsize $^{1}$ MSE: mean squared error} \\
\multicolumn{5}{l}{\scriptsize $^{2}$ Width of 95\% confidence intervals} \\
\multicolumn{5}{l}{\scriptsize $^{3}$ Empirical coverage of 95\% confidence intervals}
\end{tabular}
\end{sidewaystable*}
The performance of the imputation methods was assessed using different metrics: the mean squared error (MSE), the width of 95\% confidence intervals (CI width), and the empirical coverage of 95\% confidence intervals (Coverage). The full data scenario provides a baseline for comparison, as it describes these metrics for models fitted to the original data, without missing values. \\
In the scenario with 15\% MCAR (Scenario A), the complete-case analysis, obtained by discarding units with missing values for $W$ and then fitting a cure model on the complete data only, exhibits slightly higher MSE and wider confidence intervals compared to the full data scenario. As expected, estimates from the complete case analysis are not biased but have a larger variance since fewer observations are used. Both the exact and the approximate imputation methods performed similarly to the full data scenario, with only minor differences in CI widths, only for the estimates of the effects associated with $W$. In particular, we observe that multiple imputation lead to a decreased variance for the estimates of the parameters corresponding to fully observed variables. \\
When the missingness is increased to 30\%, still under a MCAR mechanism (Scenario B), this trend is confirmed: the complete-case analysis shows higher MSE and wider confidence intervals, while the imputation methods (via either the exact or approximate procedure) yield results close to the full data scenario. Although there are slight variations, these methods effectively preserve empirical coverage, with a slightly greater CI length when the effects of $W$ are considered. \\
These results show that the proposed approaches provide reliable parameter estimates even in the presence of a moderate to high proportion of missing data. However, the true strength of our methods becomes evident in the MAR scenario (Scenario C). Here, the proposed imputation methods significantly outperform the complete-case analysis according to all evaluation metrics. This demonstrates that when data are missing at random, both exact and approximate imputation methods can effectively mitigate the impact of missing data, producing estimates close to those obtained with full data. In contrast, the complete-case analysis performs poorly, exhibiting a marked increase in MSE and a decrease in coverage as a result of the bias. Specifically, for $\alpha_0$, the MSE of the complete-case approach increased to 0.35 and the coverage decreased to 0.32. Additionally, the CI width for $\alpha_X$ was considerably wider than that obtained when using the imputation methods. Moreover, the estimate $\beta_Z$ shows a slight deterioration compared to scenarios A and B. Indeed, when using complete-case analysis, the missing data of $W$ leads to the removal of observations that might be critical for accurately estimating $\alpha_X$ and $\beta_Z$, especially if $W$ is correlated with $X$ and $Z$. This results in biased estimates and an increased MSE. In other words, by eliminating observations with missing $W$, the model's ability to capture the true relationships involving $X$ and $Z$ may be reduced. \\
The exact conditional distribution method outperforms the approximate method in terms of CI width, indicating more precise estimates. Across all scenarios, both the exact and the approximate methods consistently provide reliable estimates, maintaining the empirical coverage of the 95\% confidence intervals. Particularly in cases where covariate $W$ is missing at random, the proposed methods significantly outperform the complete-case approach, demonstrating their ability to effectively handle complex missing data patterns.

\subsection{Comparison with a full model}
The simulation Scenario D corresponds to the method proposed by \cite{beesley2016multiple}, which we will refer to as \textit{full model} and serves as a benchmark for evaluating a further feature of the approach we propose here. In this scenario, we include in both components covariates whose impact on the outcome is null ($\alpha_Z = \beta_X = 0$). In other words, we saturate the model by including irrelevant covariates in the incidence and the latency. \\
Thus, the set of parameters can be summarized as follows: 
\begin{itemize}
    \item \textbf{Scenario D:} MAR data with 30\% missingness; \\
    $W \sim$ Normal; $\alpha_0 = 0.1$, $\alpha_W = \alpha_X = \beta_W = \beta_Z = 0.5$, and $\alpha_Z = \beta_X = 0$
\end{itemize}
The comparison between Scenarios C and D highlights the advantages of a reduced model with selected covariates in each of the two parts over the full model discussed by \cite{beesley2016multiple}. 
Focusing on the \textit{Exact} and the \textit{Approximate} methods, we observe consistent improvements in Scenario C across both the incidence and latency components. For instance, the MSE for $\alpha_W$ decreases from 0.02 in Scenario D to 0.01 in Scenario C using the exact method, and from 0.05 to 0.03 with the approximate method. This improvement is even more evident for $\alpha_X$, where the MSE decreases from 0.07 (Exact, Scenario D) to 0.04 (Exact, Scenario C), and from 0.07 to 0.05 for the approximate method.

In addition, confidence interval (CI) widths for all parameters are narrower in Scenario C. Specifically, the CI width for $\alpha_W$ improves from 0.65 (Exact, Scenario D) to 0.60 (Exact, Scenario C), and from 0.92 to 0.79 for the approximate method. A similar pattern is observed for $\beta_W$, where CI widths decrease from 0.36 (Exact, Scenario D) to 0.34 (Exact, Scenario C), and from 0.40 to 0.35 for the approximate method.

These results demonstrate the advantage of differentiating between covariates relevant to the incidence and to the latency component, as including irrelevant covariates leads to wider confidence intervals. In addition, we also observe that, if all covariates are used in both submodels when not necessary, the approximate method provides more reliable results compared to the exact approach. 
Even if the differences are modest in this simulation, since the number of parameters in the two models only differs by two, the benefits would be more substantial in scenarios with additional covariates, where the full model risks inefficiency due to overparameterization.

\subsection{Algorithm performance under model misspecification}
Appendix A and B provide the exact and approximate distributions for cases where the variable $W$ is included in only the incidence or the latency of the Cox PH cure model. However, we may lack information about the impact of $W$ on the two components. Therefore, we aim to study the performance of the proposed imputation algorithm when the model is misspecified, e.g. when $W$ is present in only one of the two components and we assume it is in both. To do this, we conduct a simulation study, examining two further scenarios: 
\begin{itemize}
    \item \textbf{Scenario E:} MAR data with 30\% missingness; \\
    $W \sim$ Normal; $\alpha_0 = 0.1$, $\alpha_W = \alpha_X = \beta_Z = 0.5$, and $\beta_W = 0$
    \item \textbf{Scenario F:} MAR data with 30\% missingness; \\
    $W \sim$ Normal; $\alpha_0 = 0.1$, $\alpha_X = \beta_W = \beta_Z = 0.5$, and $\alpha_W = 0$
\end{itemize}
These scenarios are similar to Scenario C introduced in section \ref{subsection:simul}, but they differ in that the coefficient associated with the missing covariate $W$ is set to zero in either the latency (Scenario E, $\beta_W = 0$) or the incidence component (Scenario F, $\alpha_W = 0$). This allows us to test the algorithm's ability to handle situations where the missing variable affects only one component of the Cox PH cure model. Note however, that these scenarios are not a comparison with \cite{beesley2016multiple} since the covariates $X$ and $Z$ are not the same.

Importantly, our focus is testing the algorithm's performance under the MAR scenario. We expect that if the algorithm performs well in these more complex conditions, it will likely do so in scenarios where the missing data process is either less complex or it affects a lower portion of the data. \\
This simulation study provides insights into the robustness of the mixture cure model across the two distinct scenarios. In the previous setting, we discussed the inclusion of irrelevant variables without missing values in both model components, while here we investigate the effect of including a covariate with missing values also when it does not affect one of the submodels. Ideally, we would like a multiple imputation procedure that is robust to such model misspecification. Then, the model itself, through its fit to the available data, can suggest the appropriate placement of different covariates based on the significance of the coefficient estimates. 

In Table \ref{miss}, simulation results for scenarios E and F are shown. Analogously to what reported in Table \ref{cens}, the performance of the imputation methods was assessed using the MSE, the CI width, and the empirical coverage of 95\% confidence intervals. 
\begin{table*}[h]
\caption{Simulation results under model misspecification. Evaluation metrics, $B=1000$ samples} 
\centering
 \label{miss} 
\begin{tabular}{lccccc}
\hline 
\hline \\[-1.8ex]  
\multicolumn{6}{c}{\quad \quad \quad \quad \quad \quad \textbf{MSE$^{1}$ (CI width$^{2}$) Coverage$^{3}$}} \\
 \textbf{Method} & $\alpha_{0}$ & $\alpha_{W}$ & $\alpha_{X}$ & $\beta_{W}$ & $\beta_{Z}$ \\
 \hline \\[-1.8ex] 
 \textbf{Scenario D} & & & & & \\
 \textit{Full data} & 0.03 (0.63) 0.95 & 0.01 (0.47) 0.96 & 0.05 (0.88) 0.94 & 0.00 (0.28) 0.97 & 0.02 (0.54) 0.94 \\
 \textit{Complete-case} & 0.33 (0.86) 0.31 & 0.03 (0.69) 0.96 & 0.15 (1.31) 0.93 & 0.01 (0.32) 0.95 & 0.02 (0.62) 0.95 \\
 \textit{Exact} & 0.03 (0.65) 0.94 & 0.03 (0.58) 0.91 & 0.05 (0.86) 0.94 & 0.01 (0.33) 0.98 & 0.02 (0.54) 0.94 \\
 \textit{Approximate} & 0.03 (0.69) 0.96 & 0.03 (0.87) 0.98 & 0.05 (0.92) 0.93 & 0.01 (0.34) 0.97 & 0.02 (0.54) 0.94 \\
 & & & & & \\
 \textbf{Scenario E} & & & & & \\
 \textit{Full data} & 0.02 (0.61) 0.96 & 0.01 (0.41) 0.94 & 0.04 (0.81) 0.95 & 0.01 (0.30) 0.96 & 0.02 (0.55) 0.96 \\
 \textit{Complete-case} & 0.31 (0.83) 0.33 & 0.02 (0.58) 0.95 & 0.10 (1.14) 0.91 & 0.01 (0.34) 0.95 & 0.03 (0.63) 0.94 \\
 \textit{Exact} & 0.03 (0.62) 0.95 & 0.03 (0.53) 0.92 & 0.04 (0.82) 0.95 & 0.01 (0.35) 0.90 & 0.02 (0.56) 0.95 \\
 \textit{Approximate} & 0.03 (0.70) 0.96 & 0.03 (0.72) 0.95 & 0.04 (0.83) 0.95 & 0.01 (0.36) 0.94 & 0.02 (0.58) 0.96 \\
 & & & & & \\
 \hline 
\hline \\[-1.8ex] 
\multicolumn{5}{l}{\scriptsize $^{1}$ MSE: mean squared error} \\
\multicolumn{5}{l}{\scriptsize $^{2}$ Width of 95\% confidence intervals} \\
\multicolumn{5}{l}{\scriptsize $^{3}$ Empirical coverage of 95\% confidence intervals}
\end{tabular}
\end{table*}
As before, the data without missing observations serves as a benchmark.\\ Quite similar to results in Table \ref{cens}, the complete-case analysis shows a significant increase in MSE for $\alpha_0$, reaching 0.33 in Scenario E, with a corresponding coverage of only 0.31. This result reflects the method’s vulnerability to bias when missing data is not appropriately addressed. The complete-case method also displays increased CI widths, especially for $\alpha_X$, highlighting the loss of information and precision due to missing data.

The exact method generally maintains low MSE values in both scenarios and good coverage for most parameters. However, its performance sometimes falls short of the approximate method, as seen with $\alpha_W$ in Scenario E, where the exact method yields a coverage of 0.91 compared to 0.98 of the approximate method. The higher coverage of the approximate method can be attributed to its increased CI width, which provides a more conservative estimate, thereby encompassing more potential values of the parameter. The exact method's performance again suggests it may be more sensitive to model assumptions, leading to increased bias when these assumptions are not met. Despite this, the results support conclusions about the general robustness of the model. The slightly poorer performance compared to the approximate method could be due to the tighter linkage between the missing data variable $W$ and the linear predictors, resulting in higher sensitivity to model assumptions and less flexibility in handling deviations.

The approximate method shows consistent performance with low MSEs in both scenarios and high coverage, highlighting its adaptability in dealing with model deviations. It achieves slightly better coverage for parameters like $\beta_W$ in Scenario F, reaching 0.94 compared to the exact method's 0.90.

In Scenario F, the complete-case analysis exhibits a substantial increase in MSE values for $\alpha_0$ and poor coverage, confirming its inadequacy in scenarios with high levels of missingness. Conversely, the approximate method maintains reliable coverage and CI widths across parameters due to its adaptability and reduced reliance on strict assumptions. 

Overall, these results suggest that the approximate method is particularly well-suited for real-world applications where knowledge of model structure is limited and data are affected by several sources of error. Its robust performance across scenarios E and F reinforces the idea that it is a viable alternative to the exact method.


\section{Case study}
\label{section:data}
Our method is applied to data from the MRC BO06/EORTC 80931 randomised controlled trial (RCT) for patients with localised resectable high-grade osteosarcoma. Osteosarcoma is a malignant bone tumour which mostly affects children, adolescents and young adults. Patients with osteosarcoma cannot be successfully treated with surgery alone, but adjuvant chemotherapy has been proven to significantly increase survival \citep{osteosarcoma1,osteosarcoma2}. The impact of doses and dose intensities is a topic of continuous discussion in this context \citep{intensity} and it is also relevant to other cancer types. Several previous studies have identified the presence of long-term disease-free survivors among osteosarcoma patients, who are practically considered as ‘cured’. Indeed, less than five per cent of patients experience a relapse after five or more years of follow-up. This implies that a patient falls into one of two groups once the main treatment has been administered: those who have been healed and those who have not, meaning they will eventually suffer from the progression of their disease. \\
The BO06 clinical trial includes 497 patients diagnosed between May 1993 and September 2002 \citep{trial}, and randomly assigned to receive either a conventional two-drug regimen (Reg-C), consisting of six 3-week cycles of doxorubicin (DOX, 75 mg/m2) and cisplatin (CDDP, 100 mg/m2) or a dose-intensified (DI) regimen (Reg-DI), consisting of the same doses administered twice a week and supported by G-CSF (5µg/kg daily). 

The aim of the MRC BO06/EORTC 80931 clinical trial was to investigate whether increasing the DI would improve the survival of patients with nonmetastatic limb osteosarcoma. The primary outcome measure is progression-free survival (PFS) since the end of treatment, as a cure could occur at any time during the treatment.

We excluded from the study the patients who did not receive chemotherapy, had abnormal dosages of one or both agents (more than 1.25×prescribed dose), did not have surgery, died or experienced disease progression during treatment (these last two groups were excluded because the primary outcome is progression-free survival calculated from the end of treatment), resulting in a sample of 429 individuals. \\ Some of the patients' baseline characteristics are shown in Table \ref{charact}, more details about the patients and chemotherapy can be found in the primary publication of the trial \citep{trial}. 

\begin{table}[h]
\caption{Baseline characteristics of patients in the BO06 trial.} 
\centering
 \label{charact} 
\begin{tabular}{lc}
\hline 
\hline \\[-1.8ex]  
\textbf{Characteristic} & $N = 429$ \\
 \hline \\[-1.8ex] 
 \textbf{Allocated treatment, n (\%)} & \\
 \quad  \textit{Reg-C} & 207 (48\%) \\
 \quad  \textit{Reg-DI} & 222 (52\%) \\
 \textbf{Sex, n (\%)} & \\
 \quad \textit{Female} & 173 (40 \%) \\
 \quad  \textit{Male} & 256 60\%) \\
 \textbf{Histological response, n (\%)} & \\
 \quad  \textit{Poor} & 214 (56\%) \\
 \quad  \textit{Good} & 165 (44\%) \\
 \quad  \textit{Unknown} & 50 \\
 \hline 
\hline \\[-1.8ex]  
\end{tabular}
\end{table} 

\begin{figure*}[h!]
 \centering
  \subfloat[Complete-case analysis]{\includegraphics[width=.78\linewidth]{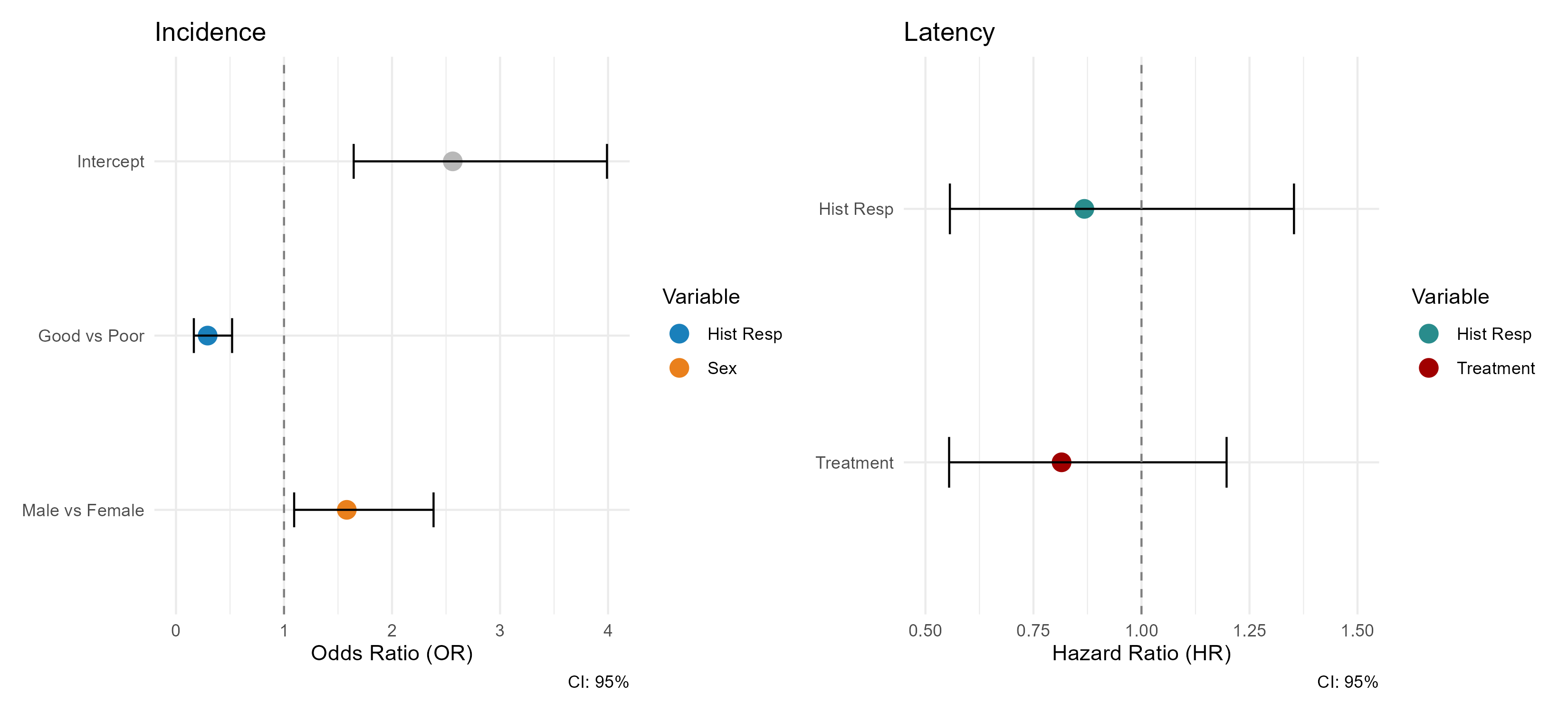}} \\
  \subfloat[Multiple imputation using the exact conditional distribution]{\includegraphics[width=.78\linewidth]{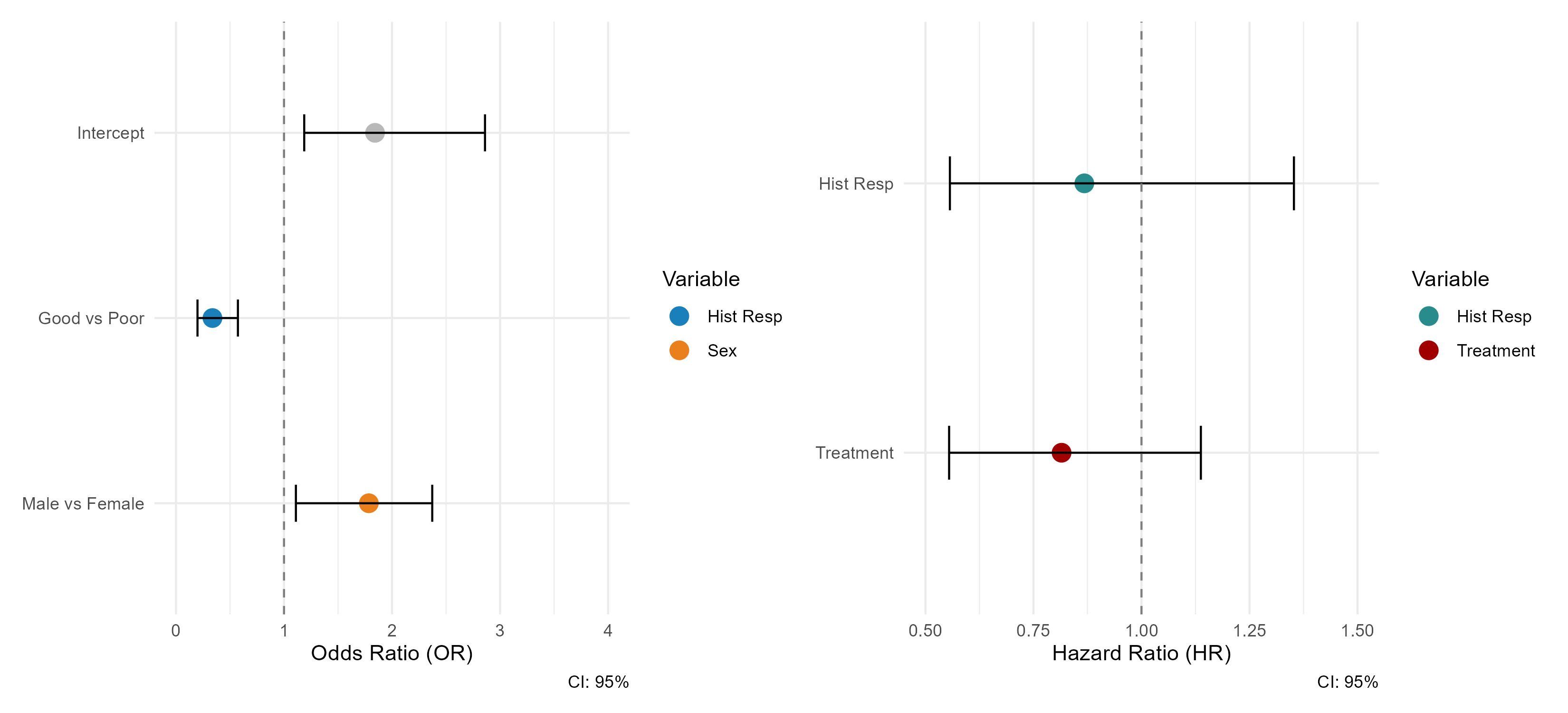}} \\
  \subfloat[Multiple imputation using the approximate conditional distribution]{\includegraphics[width=.78\linewidth]{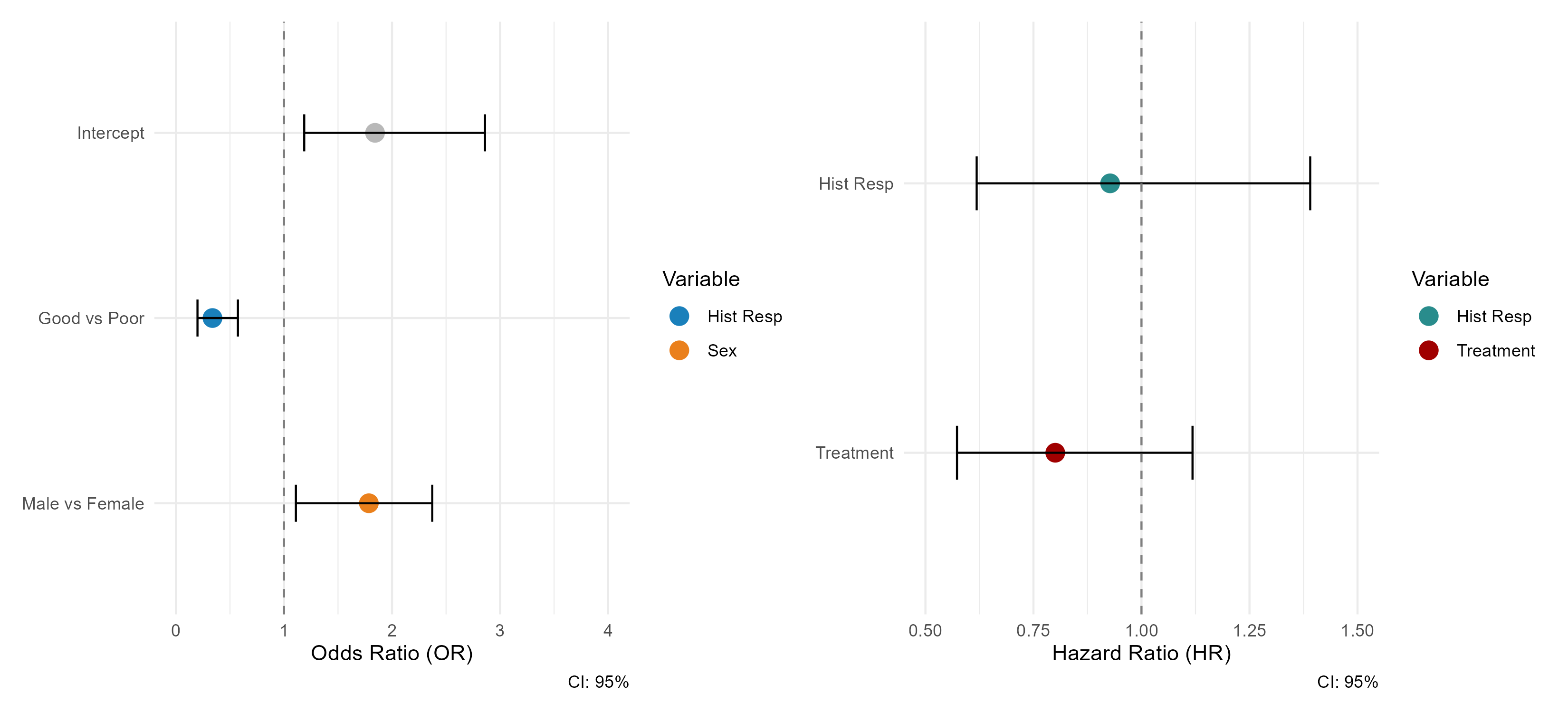}}
  \caption{\textbf{Cox proportional hazards cure model for time-to osteosarcoma recurrence.} Odds Ratios for the incidence component and Hazard Ratios for the latency component of the mixture cure models, together with 95\% confidence intervals.}
  \label{application}
\end{figure*}

Our purpose is to just perform a sensitivity analysis of the results with respect to the method used in handling the missing data. We fit a Cox proportional hazards cure model to the complete data, selecting variables for the incidence and latency components using a combination of clinical knowledge from previous studies \citep{musta2021new} and statistical considerations. We included histological response and assigned treatment as predictors in the latency component of the model. Conditioned on this choice, we selected sex and histological response for the incidence part of the model, as these variables were strongly associated with the probability of being cured. Notably, these variables would also have been selected using the penalized likelihood approach proposed by \cite{masud}. Details regarding the test for the sufficient follow-up assumption are provided in Appendix C. As histological response suffers from $\sim 12$\% missing values, we performed multiple imputation of this variable using both the exact and the approximate approaches. Figure \ref{application} illustrates the results of the Cox proportional hazards cure model for time-to-osteosarcoma recurrence. 

This figure displays the odds ratios for the incidence component and the hazard ratios for the latency component of the mixture cure models, along with 95\% confidence intervals. Panel (a) presents the results of the complete-case analysis, while panels (b) and (c) show the outcomes of multiple imputation using the exact and approximate conditional distributions, respectively. \\
Point estimates and confidence intervals are very similar between the two imputation approaches. Based on the simulation results, we do not expect big differences between the analyses based on imputed data and the complete-case one, as the percentage of missingness in the histological response variable is relatively small. \\
The only minor difference is that confidence intervals tend to be narrower for the imputation approaches than for complete-case analysis. The most notable difference between the imputation and complete-case fits is in the estimates of the model intercept, mainly in the width of his confidence interval, which is much wider in the complete-case setting. This behaviour is consistent with the results observed in the simulations.

As for the simulation in Scenario C, we also compare our extended method with the approach proposed by Beesley et al., incorporating an additional variable derived from the interaction term of histologic response and treatment. \\
Table \ref{compar} shows the results from the comparison between a full model, which includes all covariates in both the incidence and latency components, and a reduced model, obtained by selecting variables in the two parts. 

\begin{table*}[h]
\caption{Comparison of Cox PH cure model estimates with a full and a reduced model for the analysis of osteosarcoma recurrence.} 
\centering
 \label{compar} 
\begin{tabular}{lcccc|cccc}
\hline 
\hline \\[-1.8ex]  
\multicolumn{1}{l}{} & \multicolumn{4}{c|}{\textbf{Full model}} & \multicolumn{4}{c}{\textbf{Reduced model}} \\
\textbf{Complete-case} & OR$^{1}$ & 95\% CI$^{1}$ & HR$^{1}$ & 95\%  CI$^{1}$ & OR$^{1}$ & 95\% CI$^{1}$ & HR$^{1}$ & 95\%  CI$^{1}$ \\
 \hline \\[-1.8ex] 
 \textbf{Intercept} & 3.24 & 1.90, 5.50 & & & 2.56 & 1.65, 3.99 & & \\
 \textbf{Histological response} & & & & & & & & \\
 \quad  \textit{Poor} & -- & -- & -- & -- & -- & -- & -- & -- \\
 \quad  \textit{Good} & 0.19 & 0.09, 0.42 & 0.81 & 0.44, 1.51 & 0.29 & 0.17, 0.52 & 0.87 & 0.56, 1.35 \\
 \textbf{Sex} & & & & & & & & \\
 \quad \textit{Female} & -- & -- & -- & -- & -- & -- & & \\
 \quad  \textit{Male} & 1.48 & 1.01, 2.18 & 1.16 & 0.87, 1.54 & 1.58 & 1.09, 2.28 & & \\
 \textbf{Allocated treatment} & & & & & & & & \\
 \quad  \textit{Reg-C} & -- & -- & -- & -- & & & -- & -- \\
 \quad  \textit{Reg-DI} & 0.53 & 0.25, 1.10 & 0.90 & 0.57, 1.42 & & & 0.81 & 0.55, 1.20 \\
 \textbf{Treatment*Hist. resp.} & & & & & & & & \\
 \quad \textit{Good.Reg-DI} & 2.67 & 0.92, 7.74 & 0.99 & 0.42, 2.31 & & & &  \\
 \midrule 
\textbf{Exact} & & & & & & & & \\
\midrule \\[-1.8ex]  
 \textbf{Intercept} & 2.45 & 1.37, 4.41 & & & 1.85 & 1.14, 3.00 & & \\
 \textbf{Histological response} & & & & & & & & \\
 \quad  \textit{Poor} & -- & -- & -- & -- & -- & -- & -- & -- \\
 \quad  \textit{Good} & 0.19 & 0.09, 0.42 & 0.81 & 0.44, 1.52 & 0.29 & 0.17, 0.52 & 0.87 & 0.56, 1.35 \\
 \textbf{Sex} & & & & & & & & \\
 \quad \textit{Female} & -- & -- & -- & -- & -- & -- & & \\
 \quad  \textit{Male} & 1.74 & 1.01, 3.00 & 1.24 & 0.84, 1.82 & 1.91 & 1.14, 2.22 & & \\
 \textbf{Allocated treatment} & & & & & & & & \\
 \quad  \textit{Reg-C} & -- & -- & -- & -- & & & -- & -- \\
 \quad  \textit{Reg-DI} & 0.53 & 0.25, 1.10 & 0.90 & 0.57, 1.42 & & & 0.82 & 0.55, 1.14 \\
 \textbf{Treatment*Hist. resp.} & & & & & & & & \\
 \quad \textit{Good.Reg-DI} & 2.67 & 0.92, 7.75 & 0.99 & 0.42, 2.31 & & & &  \\
 \midrule
 \textbf{Approximate} & & & & & & & & \\
 \hline \\[-1.8ex] 
 \textbf{Intercept} & 2.12 & 1.24, 3.61 & & & 1.84 & 1.19, 2.86 & & \\
 \textbf{Histological response} & & & & & & & & \\
 \quad  \textit{Poor} & -- & -- & -- & -- & -- & -- & -- & -- \\
 \quad  \textit{Good} & 0.24 & 0.11, 0.49 & 0.90 & 0.50, 1.62 & 0.34 & 0.20, 0.57 & 0.93 & 0.62, 1.39 \\
 \textbf{Sex} & & & & & & & & \\
 \quad \textit{Female} & -- & -- & -- & -- & -- & -- & & \\
 \quad  \textit{Male} & 1.68 & 1.01, 2.80 & 1.18 & 0.82, 1.70 & 1.79 & 1.11, 2.37 & & \\
 \textbf{Allocated treatment} & & & & & & & & \\
 \quad  \textit{Reg-C} & -- & -- & -- & -- & & & -- & -- \\
 \quad  \textit{Reg-DI} & 0.75 & 0.37, 1.53 & 0.82 & 0.53, 1.26 & & & 0.80 & 0.57, 1.12 \\
 \textbf{Treatment*Hist. resp.} & & & & & & & & \\
 \quad \textit{Good.Reg-DI} & 2.12 & 0.75, 6.03 & 0.95 & 0.43, 2.11 & & & &  \\
 \hline 
\hline \\[-1.8ex]  
\multicolumn{9}{l}{$^{1}$ OR: Odds Ratio, CI: Confidence Interval, HR: Hazard Ratio}
\end{tabular}
\end{table*} 

The estimates from the saturated model show a wider uncertainty in parameter estimates. This is evident in the broader confidence intervals, especially for the intercept, sex, and treatment effects, indicating that the inclusion of unnecessary covariates introduces additional variability and reduces the precision of the estimates.
In contrast, for the non-saturated models, both Exact and Approximate, the confidence intervals are narrower. 
The odds ratios for "Good" histological response are consistent across both imputation methods, but the confidence intervals are tighter in the Exact method, as observed in the simulation study.
Overall, we observe that the precision of the parameter estimates improves when we limit the covariates to those that are meaningful in the context of the disease and treatment under study. This suggests that distinguishing between the covariates affecting the probability of being cured and those influencing the survival of uncured patients may be crucial for more efficient modelling.

\section{Discussion}
\label{section:discussion}
Our work on imputation methods for mixture cure models in the presence of missing covariate values emphasises the importance of flexible imputation techniques. In this paper, we investigate the methodology for imputing partially observed covariates in a mixture cure model by accommodating potentially distinct sets of covariates for the cure probability and the survival of uncured patients. This extension is motivated by both theoretical considerations - such as the identifiability benefits of distinct covariate sets - and practical challenges that arise when including all available covariates to both submodels. Importantly, we release the \texttt{R}-code to implement the proposed approach. 

The results show that both exact and approximate imputation approaches produce similar results when the model is correctly specified. In particular, imputation methods yield narrower confidence intervals compared to the complete case analysis, enhancing the precision of our estimates. 
The simulation results suggest that the proposed methods are robust to the presence of missing values, particularly under the MAR scenario, where they outperform the complete case analysis. This highlights the effectiveness of multiple imputation in handling complex missing data patterns, preserving empirical coverage, and reducing MSE. The exact imputation method slightly outperforms the approximate method in terms of confidence interval width, yielding more precise estimates. However, when the model is misspecified, i.e. certain covariates are wrongly included in the incidence or latency components, the approximate method has higher coverage probabilities, hence it is a safer and more robust choice for practical applications. Furthermore, the results illustrate the advantage of differentiating covariates between incidence and latency, since wrongly including all covariates in both components leads to wider confidence intervals and worse performance of the exact approach.

For data from the BO06 clinical trial on osteosarcoma, the application of the Cox proportional hazards cure model reveals several key findings. First, as observed in the complete case analysis, histological response remains a strong prognostic factor for the cure fraction, indicating that patients who respond well to initial treatment are more likely to be cured. However, even after accounting for missing values, there is no clear indication that a good histological response is associated with longer progression-free survival (PFS) for uncured patients. This underscores the importance of having the possibility to differentiate between factors that influence the probability of cure from those that affect survival among uncured patients. Overall, we observe that, despite slight changes in the estimated coefficients and of the length of the confidence intervals, the conclusions about the significance of the covariate effects are the same for all the models and estimation procedures considered.

As a further development of this work, one might consider an alternative way of performing multiple imputation based on the observed likelihood (instead of the complete one), which avoids considering the latent cure status as an additional variable with missing values. When there is only one missing variable, this would avoid the chained equations, since there would be no need to impute G. However, such an approach would require the estimation of the parameters $\alpha$ and $\beta$ in Step 2 of Algorithm \ref{alg:cap} by fitting a mixture cure model and the variance needs to be estimated via computationally intensive bootstrap. Hence, we will explore this possibility in future works.

\section*{Supplementary material}
The source codes for implementing the imputation algorithm and for reproducing the simulations are available at \url{https://github.com/martacip/mi_curemodels}.

\section*{Acknowledgments}

MRC Clinical Trials Unit at UCL (London, UK) and the European Organisation for Research and Treatment of Cancer (EORTC) are gratefully acknowledged for making data from the MRC BO06/EORTC 80931 trial available for this analysis.
The trial was supported and coordinated by the U.K. MRC and the EORTC.
\bibliographystyle{plainnat}
\bibliography{article}

\section*{Appendix A - Imputation using the exact conditional distribution}
We derive an exact imputation model for imputing both normal and binary covariates in the case in which the partially observed covariate belongs either to the incidence component or to the latency component of the Cox PH mixture cure model. 

\subsection*{A.1 Incomplete covariate only in the incidence}
\textit{Case 1.} Suppose $W \vert X^{(-j)}, Z \sim Normal(\mu, \sigma)$ where
\begin{equation}
\label{mu}
    \mu = \theta_0 + \sum\limits_{b \neq j}^{p} \theta_b X_b + \sum\limits_{\substack {s \\ Z_s \notin X}}^{d}{\theta_{p+s}} Z_s
\end{equation}
Updating the complete data likelihood for the Cox PH cure model given in \eqref{eqn:L_Norm}, we obtain the following exact conditional distribution for $W$ 
\begin{align}
\begin{split}
\label{eqn:fW_Norm3}
 f(W_i \vert X_{i}^{(-j)}, Z_{i}, Y_i, \Delta_i, G_i) 
 \propto \left( \frac{e^{\alpha_0+\alpha^TX_i}}{1+e^{\alpha_0+\alpha^TX_i}}
\right)^{G_i}
\left( \frac{1}{1+e^{\alpha_0+\alpha^TX_i}}
\right)^{1-G_i}
\left( e^\frac{-(W_i-\mu_i)^2}{2\sigma^2} \right)
\end{split}
\end{align}
\textit{Case 2.} Let assume that $W \vert X^{(-j)}, Z \sim Bernoulli(\expit(\mu))$, where $\mu$ is given by \eqref{mu}, then the exact distribution of $W$ takes the form
\begin{equation}
    \logit \bigl[ P(W_{i} = 1 \vert X_{i}^{(-j)}, Z_{i}, Y_{i}, \Delta_{i}, G_{i}) \bigr] = G_{i}\alpha_{j} + \log\bigl(1 + e^{\alpha_{0} + \sum_{b \neq j}^{p}{\alpha_b X_{i,b}}} \bigr) - \log\bigl( 1 + e^{\alpha_{0} + \alpha_{j} + \sum_{b \neq j}^{p}{\alpha_b X_{i,b}}} \bigr) + \mu_{i}
\end{equation}

\subsection*{A.2 Incomplete covariate only in the latency}
\textit{Case 1.} If we assume that $W \vert X, Z^{(-l)} \sim \text{Normal}(\mu, \sigma)$ and we define $\mu$ as
\begin{equation}
    \label{mu2}
    \mu = \theta_0 + \sum\limits_{b }^{p} \theta_b X_b + \sum\limits_{\substack {s \neq l \\ Z_s \notin X}}^{d}{\theta_{p+s}} Z_s
\end{equation}
we can derive the following exact conditional distribution for $W$
\begin{align}
\begin{split}
\label{eqn:fW_Norm2}
 f(W_i \vert X_{i}, Z_{i}^{(-l)}, Y_i, \Delta_i, G_i) 
 \propto \left( e^{\beta^T Z_i \Delta_i}
e^{-H_0(Y_i) e^{\beta^T Z_i}}
\right)^{G_i}
\left( e^\frac{-(W_i-\mu_i)^2}{2\sigma^2} \right)
\end{split}
\end{align}
\textit{Case 2.} Suppose that $W \vert X, Z^{(-l)} \sim \text{Bernoulli}(\expit(\mu))$, where $\mu$ is as in \eqref{mu2}, then $W$ follows the distribution
\begin{equation}
\label{last}
    \logit \bigl[ P(W_{i} = 1 \vert X_{i}, Z_{i}^{(-l)}, Y_{i}, \Delta_{i}, G_{i}) \bigr] = G_i \Delta_i \beta_l
- G_i H_0(Y_i) e^{\sum_{s \neq l}^{d}{\beta_s Z_{i,s}}} (e^{\beta_l}-1) + \mu_{i}
\end{equation}

\section*{Appendix B - Imputation using the approximate conditional distribution}
Similarly to what was discussed in Section 3.3 and by utilizing approximations \eqref{eqn:approx1} and \eqref{eqn:approx2}, we can derive an approximate version of the exact distributions shown in Appendix A, which allows us to impute the missing values of W at a reduced computational cost.

\subsection*{B.1 Incomplete covariate only in the incidence}
\textit{Case 1.} We assume that $W \vert X^{(-j)}, Z \sim Normal(\mu, \sigma)$ and that $\mu$ is equal to \ref{mu}. An approximation of the exact conditional distribution in \ref{eqn:fW_Norm3} is given by 
\begin{equation}
\log \left( f(W_i \vert X_{i}^{(-j)}, Z_{i}, Y_i, \Delta_i, G_i) \right)
\approx -\frac{1}{2\sigma^{2}}W_{i}^{2} + \left[G_i \alpha_j 
- \frac{e^{\alpha_0+\alpha^{T}\bar{X}}}{1+e^{\alpha_0 + \alpha^T \bar{X}}} \alpha_j 
+ \frac{1}{\sigma^2} \mu_i
\right] W_i
+C
\end{equation}
where $C$ is a constant term. A linear regression function with the covariates $X_{i}^{(-j)}$, $Z_{i}^{(-l)}$, and $G_{i}$ can be used to approximate the exact distribution of $W$.\\
\textit{Case 2.} If $W \vert X^{(-j)}, Z$ is Bernoulli distributed with parameter $\expit(\mu)$, with $\mu$ equal to \ref{mu}, an approximate imputation model is as follows
\begin{align}
\logit \left[ P(W_i=1 \vert X_{i}^{(-j)}, Z_{i}, Y_i, \Delta_i, G_i) \right]
& \approx G_i \alpha_j
+ \frac{e^{\alpha_0+\sum_{b \neq j}^{p}{\alpha_b \bar{X}_b}}}{1+e^{\alpha_0+\sum_{b \neq j}^{p}{\alpha_b \bar{X}_b}}} 
\left[ 1 + \sum_{b \neq j}^{p}{\alpha_b \left( X_{i,b} - \bar{X}_{b} \right)} \right] + \notag\\
& - \frac{e^{\alpha_0+\alpha_j+\sum_{b \neq j}^{p}{\alpha_b \bar{X}_b}}}{1+e^{\alpha_0+\alpha_j+\sum_{b \neq j}^{p}{\alpha_b \bar{X}_b}}} 
\left[ 1 + \sum_{b \neq j}^{p}{\alpha_b \left( X_{i,b} - \bar{X}_{b} \right)} \right]
+ \mu_i + C
\end{align}
A logistic regression function with the covariates $X_{i}^{(-j)}$, $Z_{i}^{(-l)}$, and $G_{i}$ can be used to approximate the exact distribution of $W$.

\subsection*{B.2 Incomplete covariate only in the latency}
\textit{Case 1.} We can approximate the above equation \eqref{eqn:fW_Norm2} by
\begin{equation}
\log \left( f(W_i \vert X_{i}, Z_{i}^{(-l)}, Y_i, \Delta_i, G_i) \right)
\approx - \frac{1}{2 \sigma^2} W_i^2 + \left[ G_i \Delta_i \beta_l 
- G_i H_0(Y_i) e^{\beta^{T}\bar{Z}} \beta_{l} 
+ \frac{1}{\sigma^2} \mu_i
\right] W_i
+C
\end{equation}
A linear regression function with the covariates $X_{i}^{(-j)}$, $Z_{i}^{(-l)}$, $G_{i} \Delta_{i}$, and $G_{i}H_{0}(Y_{i})$ can be used to approximate the exact distribution of $W$. \\
\textit{Case 2.} A simpler approximated model for \eqref{last} is given by 
\begin{align}
\logit \left[ P(W_i=1 \vert X_{i}, Z_{i}^{(-l)}, Y_i, \Delta_i, G_i) \right]
& \approx G_i \Delta_i \beta_l
- G_i H_0(Y_i) (e^{\beta_l}-1) e^{\sum_{s \neq l}^{d}{\beta_s \bar{Z}_{s}}}
\left[ 1 + \sum_{s \neq l}^{d}{\beta_s \left( Z_{i,s} - \bar{Z}_{s} \right)} \right] \notag \\ 
& \qquad+ \mu_i + C
\end{align}
A logistic regression function with the covariates $X_{i}^{(-j)}$, $Z_{i}^{(-l)}$, $G_{i} \Delta_{i}$, $G_{i}H_{0}(Y_{i})$, and $G_{i}H_{0}(Y_{i})Z_{i}^{(-l)}$ can be used to approximate the exact distribution of $W$.

\section*{Appendix C - Testing the sufficient follow-up assumption for the osteosarcoma data}

We test the sufficient follow-up assumption for the osteosarcoma data analyzed in Section \ref{section:data}. Currently there exist no test uniformly over all covariate values so we apply the tests proposed by \cite{assump1} and \cite{assump2} separately for patients with good and poor histologic response, since that is the main prognostic factor for the cure probability. 

The test proposed by \cite{assump1} considers the null hypothesis of sufficient follow-up $H_0: \tau_{F_u}\leq\tau_C$, where $\tau_{F_u}$ and $\tau_C$ denote, respectively, the right end points of the distribution of the event times for the uncured patients and of the censoring times. For patients with good histologic response, such test does not reject the null hypothesis of sufficient follow-up since $2(Y_{(n)}-\tilde{Y}_{(m)})<Y_{(n)}$, where $Y_{(n)}$ denotes the largest observed time and $\tilde{Y}_{(m)}$ denotes the largest observed event time. For patients with poor histologic response, the p-value of the test is 0.038 meaning that the null hypothesis is rejected at level 0.05.  

However, it is argued by \cite{assump2} that one could instead test a more relaxed notion of practically sufficient follow-up. The null hypothesis is that the follow-up is insufficient $\tilde{H}_0: \tau_C\leq q_{1-\epsilon}$, where $q_{1-\epsilon}$ denotes the $(1-\epsilon)$-quantile of $F_u$ and $\epsilon$ is a small number chosen by the user. It essentially tests whether there is less that $\epsilon$ probability for the event to happen after the end of the study. In practice this would be considered as a sufficiently long follow-up if $\epsilon$ is around $1$-$2\%$. The test requires also the choice of a parameter $\tau$ such that the chances of the event happening after $\tau$ are negligible, i.e. a study of length $\tau$ would have been certainly sufficient. We apply the test for two different choices of $\epsilon$ and $\tau$ and p-values are given in Table~\ref{tab:test}. The results indicate that the follow-up is clearly sufficient for patients with good histologic response and borderline sufficient for those with poor histologic response. 
\begin{table}[h!]
    \centering
    \begin{tabular}{c|cc|cc}
    &\multicolumn{2}{c|}{good hist.}&\multicolumn{2}{c}{poor hist.}\\
    & \multicolumn{2}{c|}{$\tau$} &\multicolumn{2}{c}{$\tau$}\\ 
       $\epsilon$  & 15 years & 20 years & 15 years & 20 years\\
       \hline
       0.01  & 0.000 & 0.000& 0.095& 0.131\\
       0.02 & 0.000&0.000&0.046&0.093
    \end{tabular}
    \caption{P-values for the testing the null hypothesis $\tilde{H}_0$ of practically sufficient follow-up for patients with good and poor histologic response.}
    \label{tab:test}
\end{table}

\end{document}